\documentclass{PoS}

\def\bea{\begin{eqnarray}}
\def\eea{\end{eqnarray}}

\title{Parton fragmentation in nuclear collisions}

\ShortTitle{Parton fragmentation in nuclear collisions}

\author{\speaker{Thomas A.~Trainor}
\\
        CENPA 354290, University of Washington, Seattle, WA 98195\\
        E-mail: \email{trainor@hausdorf.npl.washington.edu}}

\abstract{The hydrodynamic (hydro) model has been extensively applied to heavy ion data from the relativistic heavy ion collider (RHIC). Results are interpreted to conclude that a dense QCD medium nearly opaque to most partons, a strongly-coupled quark-gluon plasma (sQGP), is formed in more-central Au-Au collisions. The sQGP may have a very small viscosity (``perfect liquid''). 

However, other analysis methods provide contradictory evidence. 
Two-component analysis of single-particle hadron spectra reveals a spectrum hard component consistent with a parton fragment distribution described by pQCD which can masquerade as ``radial flow'' in some hydro-motivated analysis. Minimum-bias angular correlations reveal that a large number of back-to-back jets from initial-state scattered partons with energies as low as 3 GeV survive as ``minijet'' hadron correlations even in central Au-Au collisions, suggesting near transparency to partons. 

In this talk I present methods by which almost all spectrum and correlation structure, even in the most-central Au-Au collisions at 200 GeV, can be described quantitatively by pQCD calculations. The evolution of nuclear collisions is apparently dominated by parton scattering and fragmentation even in the most-central A-A collisions, albeit the fragmentation process is strongly modified.

}

\FullConference{Workshop on Critical Examination of RHIC Paradigms - CERP2010\\
		April 14-17, 2010\\
		Austin, Texas USA}

\begin{document}

 \section{Introduction}


The intention of this work is to test the extent to which perturbative QCD (pQCD) can describe more-central A-A collisions at RHIC. Is a hydrodynamic (hydro) description necessary, or even allowed by data? Detailed arguments are provided in Refs.~\cite{hardspec,fragevo,tzyam,nohydro}, with related material on hydro interpretations of azimuth quadrupole structure in Refs.~\cite{newflow,gluequad,quadspec,kettler}.

I begin with the two-component spectrum model for p-p collisions. Then I review the phenomenology of fragmentation functions (FFs) from LEP, HERA and Fermilab. I describe calculations on that basis of perturbative QCD (pQCD) fragment distributions (FDs). What was described this morning as ``fragmentation functions'' I distinguish from true FFs which are distributions {conditional} on parton energy. Fragment distributions can be calculated by folding FFs with a parton spectrum. I introduce a parton ``energy-loss'' model~\cite{borg} to provide FD calculations which can describe measured fragmentation evolution with A-A collision centrality. The last part of the talk extrapolates beyond the single-particle system to describe a method for converting jet angular correlations into fragment yields and spectra. It is then possible to determine quantitatively the minijet contribution to the A-A final state. What emerges is a comprehensive pQCD description of RHIC nuclear collisions up to central Au-Au.

\section{Visualizing fragmentation: Conceptual consequences of plotting formats}

\begin{figure*}[h]
  \includegraphics[width=2.9in,height=1.65in]{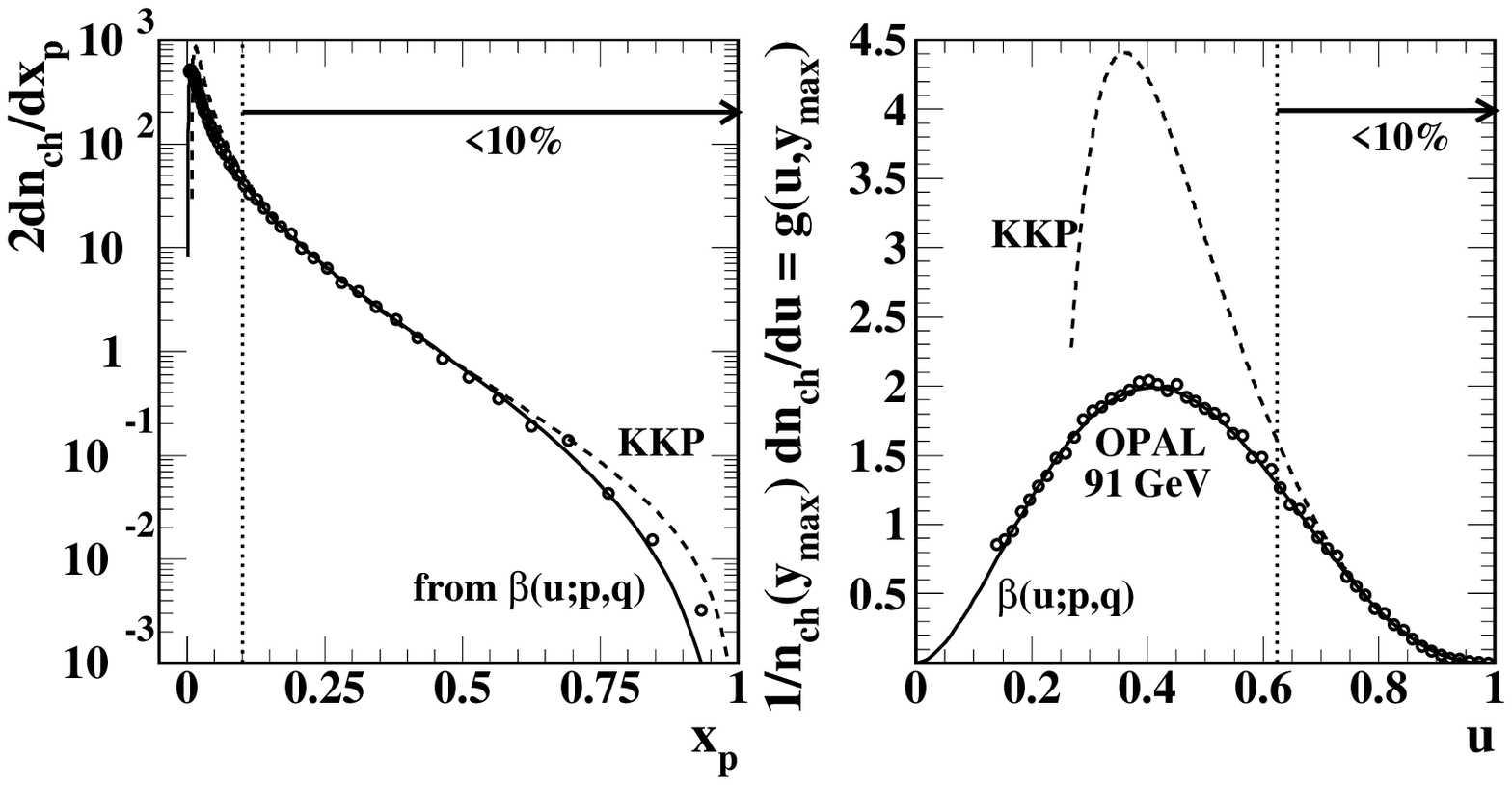}
  \includegraphics[width=1.45in,height=1.65in]{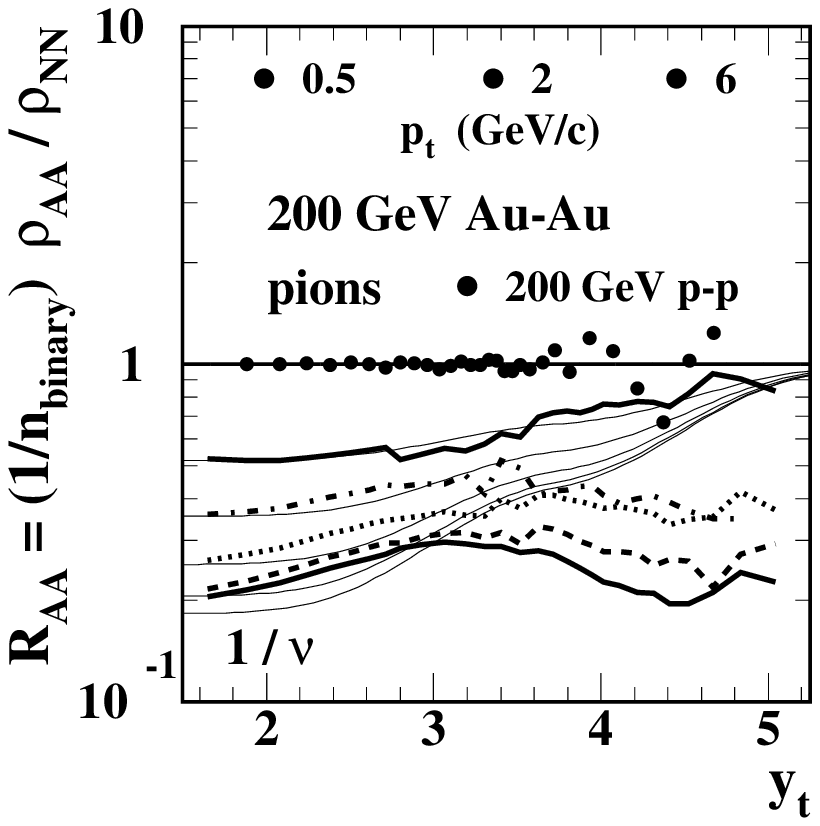}
  \includegraphics[width=1.45in,height=1.65in]{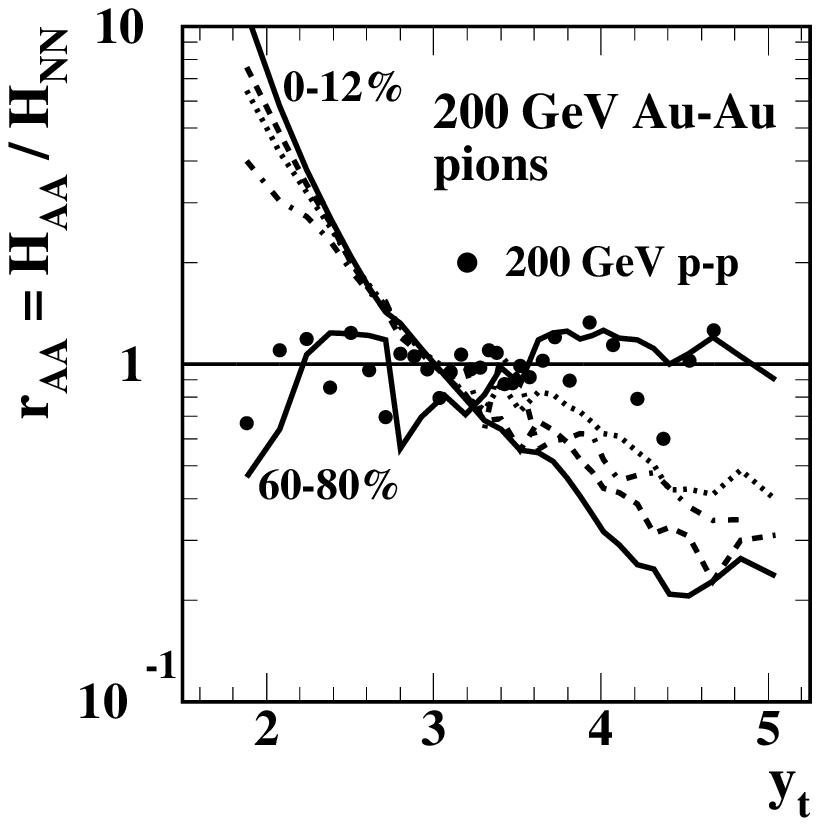}
\caption{\label{format} First: Fragmentation function for 91 GeV $e^+$-$e^-$ collisions~\cite{opal} plotted on conventional variable $x_p$. Second: The same data plotted on normalized rapidity $u$. The solid curve is a beta distribution. Third: Conventional spectrum ratio $R_{AA}$ for five centralities of 200 GeV Au-Au collisions (bold curves) and p-p data (solid dots). The thin solid curves are linear superposition references. Fourth: The same data plotted in the form of hard-component ratio $r_{AA}$, revealing large enhancements at smaller $p_t$ corresponding to suppression (jet quenching) at larger $p_t$.
 } 
 \end{figure*}

I  first consider the impact of plotting-format choices on physical interpretations of data: For instance, to what extent does a given plotting format favor hydro over fragmentation interpretations?  Figure~\ref{format} (first panel) shows a conventional plotting format for fragmentation functions, beautiful LEP data from OPAL at 91 GeV that Yuri showed  you. In this format the featured structure at larger $x_p$ actually represents a small fraction of the fragment yield---less than 10\% of the fragments---that which can be described by DGLAP evolution. Some people are quite interested in those details. The structure at upper left in the first panel (small $x_p$) is typically ignored at RHIC. 

  Figure~\ref{format} (second panel) shows the same data replotted on normalized rapidity $u$: rapidity variable $y = \ln\{(E+p)/m_\pi\}$ divided by the equivalent measure $y_{max} = \ln(Q/m_\pi)$ for the parton. We find that FF data so plotted can be described by a {\em beta} distribution (solid curve) to the error limits of the data down to zero momentum~\cite{ffprd}. 
The solid curve is not a theoretical description, it is a phenomenological description of measured FFs. DGLAP applies toward the right and MLLA applies near the peak, but the beta distribution accurately describes all fragment data from parton momentum down to zero momentum (in both panels).

 Figure~\ref{format} (third panel) shows conventional spectrum ratio $R_{AA}$ for five centralities of 200 GeV Au-Au collisions plotted on transverse rapidity $y_t = \ln\{(m_t + p_t) / m_\pi\}$~\cite{hardspec} and p-p data from Ref.~\cite{ppprd}.  $R_{AA}$ is defined as the ratio of an A-A $p_t$ spectrum to a reference p-p spectrum divided by the (Glauber) number of binary collisions. The p-p (N-N) reference $\rho_{NN}$ is in this case the two-component model described in the next section. The conventional emphasis is on $p_t$ above 6 GeV/c (upper scale) which is 4.5 on  $y_t$ (lower scale). Information about fragmentation below 6 GeV/c is strongly suppressed by $R_{AA}$, presenting a misleading picture. 

Alternatively, one can extract from the same spectra ``hard components'' $H_{AA}$, which are most relevant to fragmentation, and replot them as ratio $r_{AA}$~\cite{hardspec,ppprd}. In Fig,~\ref{format} (fourth panel) $r_{AA}$ reveals for the first time that hard-component variations at 0.5 GeV/c (large enhancement) are exactly complementary to variations at 10 GeV/c (suppression). Centrality evolution in the two places is strongly correlated. We should not claim an understanding of fragmentation (e.g. jet quenching) until the entire fragmentation picture is acknowledged.





 \section{Two-component model of spectra and correlations}


The two-component model has several manifestations at RHIC. I refer to the two-component spectrum model as first developed in Ref.~\cite{ppprd}.
The basic physical model is similar to that in PYTHIA, but the details are determined by data phenomenology. In p-p collisions the ``soft component'' refers to longitudinal fragmentation of projectile nucleons by soft-Pomeron exchange leading to diffractive dissociation. The ``hard component'' refers to large-angle scattered parton fragmentation from a minimum-bias parton spectrum, possibly by hard-Pomeron exchange. The fragment hadron distribution extends in principle down to zero momentum.

\begin{figure*}[h]
  \includegraphics[width=1.45in,height=1.65in]{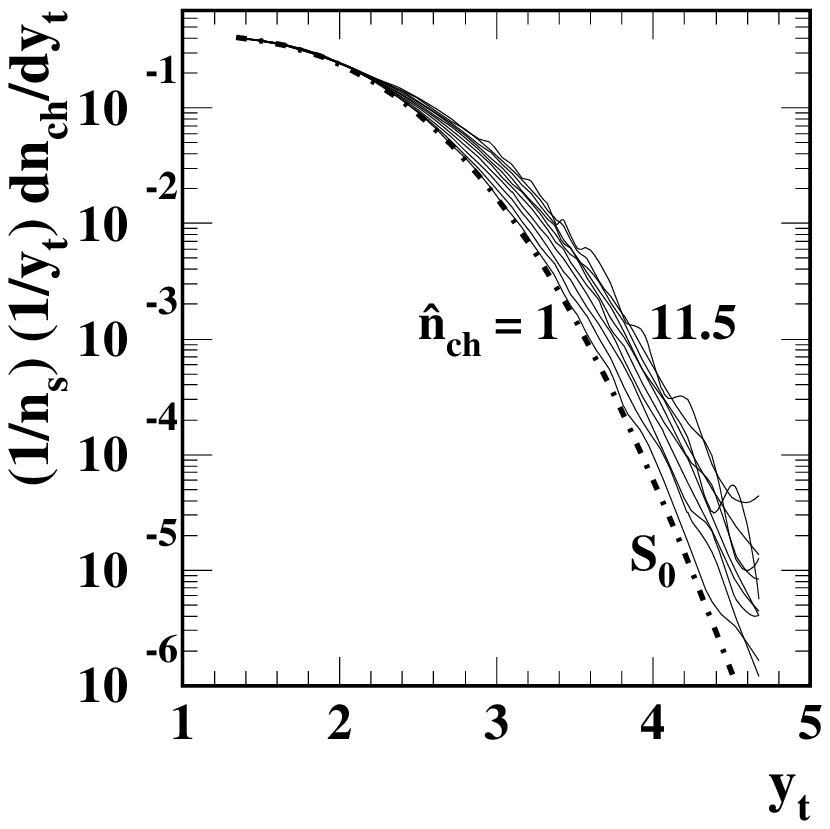}
  \includegraphics[width=1.45in,height=1.65in]{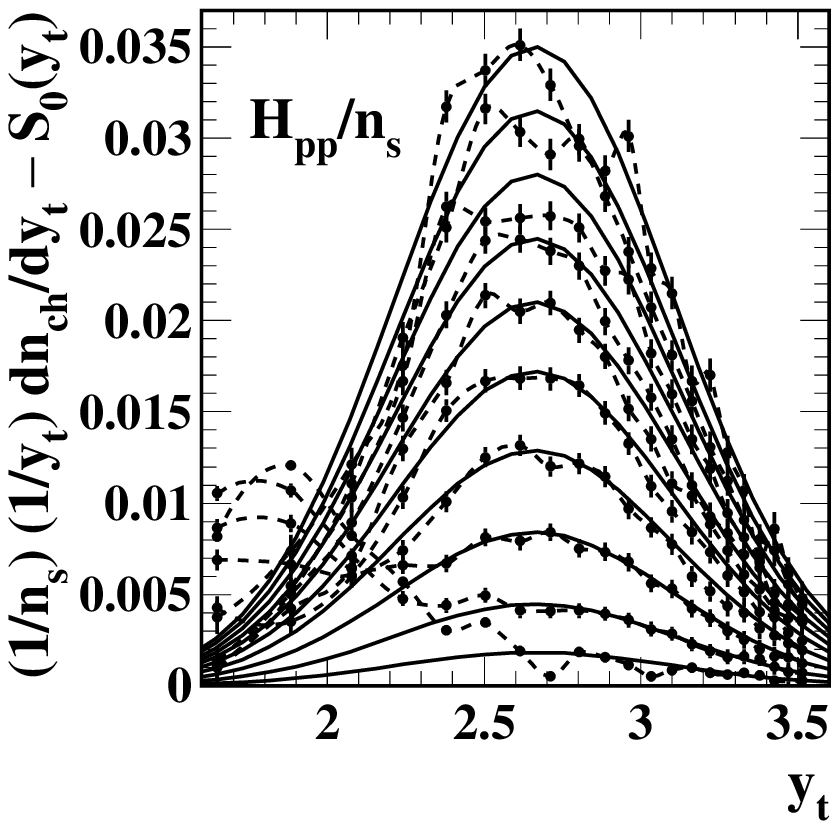}
  \includegraphics[width=1.45in,height=1.65in]{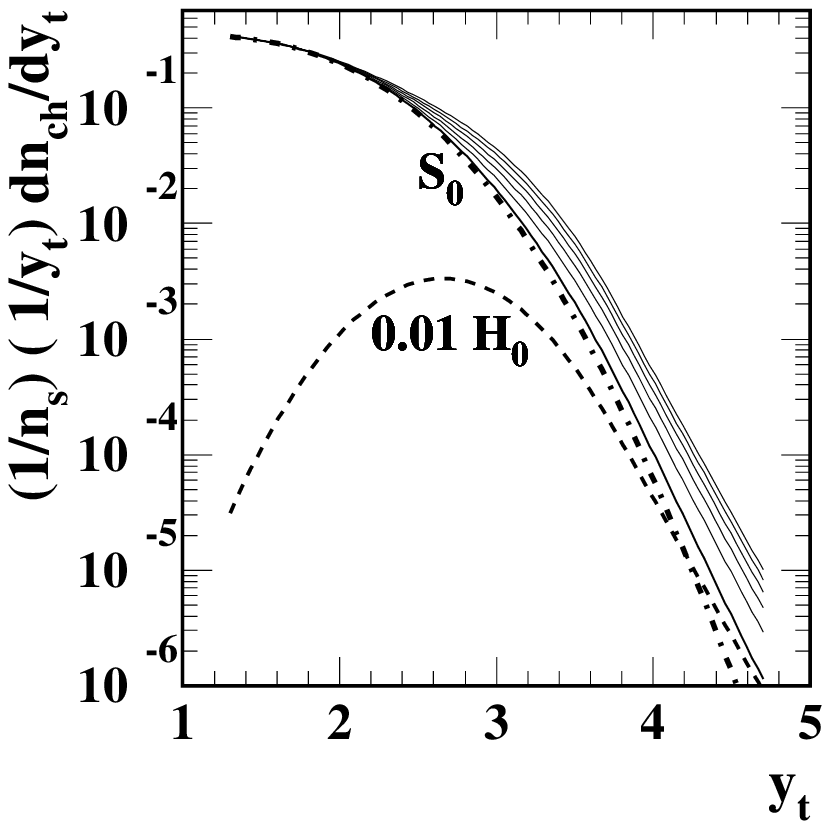}
  \includegraphics[width=1.45in,height=1.65in]{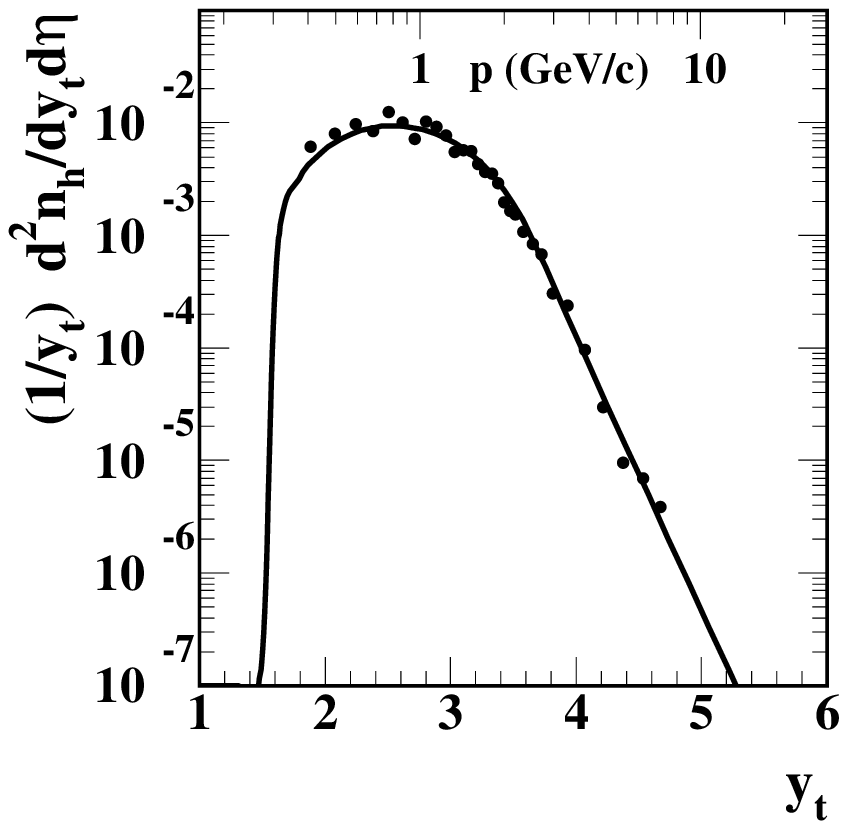}
\caption{\label{2comp} First: Spectra for ten multiplicity classes of 200 GeV NSD p-p collisions normalized to soft-component multiplicity $n_s$~\cite{ppprd}. $S_0(y_t)$ is the ``soft component'' limiting case for $n_{ch} \rightarrow 0$. Second: Spectrum hard components inferred from the same spectra by subtracting the soft component from all spectra. Third: Two-component model of the p-p spectra. $H_0(y_t)$ is the hard-component model function. Fourth: Average hard component for NSD p-p collisions (data points) with pQCD ``fragment distribution'' (solid curve).
 } 
 \end{figure*}

Figure \ref{2comp} (first panel) shows $p_t$ spectra from non-single-diffractive (NSD) p-p collisions at $\sqrt{s} = 200$ GeV~\cite{ppprd}. The spectra correspond to ten observed (uncorrected) multiplicities $\hat n_{ch}$ in one unit of $\eta$. 
Corrected multiplicity $n_{ch} \approx 2\hat n_{ch}$. 
The spectra have been normalized by ``soft'' multiplicity $n_s$ determined iteratively by a limit process. The data are plotted on transverse rapidity $y_t$ with pion mass representing unidentified hadrons. The spectra are then functions of $y_t$ and $\hat n_{ch}$. 

Evolution with $\hat n_{ch}$ is  simple. The spectra can be represented accurately by a Taylor expansion in $\hat n_{ch}$ which has only two terms. The ``constant'' term $S_0$ is a fixed function of $y_t$, and the ``linear coefficient'' $H_0$ is also a fixed function of $y_t$, both independent of $\hat n_{ch}$.
The two-component model for p-p collisions with soft and hard multiplicities $n_s + n_h = n_{ch}$ is then
\bea \label{twoc}
 \frac{1}{n_s(\hat n_{ch})}\frac{1}{y_t}\, \frac{dn_{ch}(\hat n_{ch})}{dy_t } =  S_0(y_t)  +  \frac{n_h(\hat n_{ch})}{n_s(\hat n_{ch})}\,  H_{0}(y_t),
\eea
Factor $n_h  / n_s$ is observed to vary as $\alpha\, \hat n_{ch}$.
$S_0(y_t)$ is by definition the limiting spectrum as $\hat n_{ch} \rightarrow 0$ and has the form of a L\'evy distribution on $m_t$. By subtracting $S_0(y_t)$ from each of the spectra in the first panel we obtain the residuals in the second panel. The form is independent of multiplicity and well described by the solid curves representing fixed form $H_0(y_t)$, a Gaussian plus QCD power-law tail on transverse rapidity $y_t$~\cite{ppprd}. The third panel shows the model in Eq.~(\ref{twoc}) which can be compared with data in the first panel.
For comparisons with A-A spectra (below) we define $S_{pp} =(1/y_t)\, dn_s/dy_t$ with reference model $n_s\, S_0$ and similarly for $H_{pp} \leftrightarrow n_h\, H_0$.  

In Fig.~\ref{2comp} (fourth panel) the points (spectrum hard component) represent an average of the \mbox{p-p} hard components in the second panel, each scaled to the multiplicity density corresponding to NSD \mbox{p-p} collisions. The solid curve represents a calculated pQCD {fragment distribution} discussed below.


The corresponding two-component model for per-participant-pair A-A spectra is 
\bea  \label{aa2comp}
\frac{2}{n_{part}} \frac{1}{y_t}\frac{dn_{ch}}{dy_t} &=& S_{NN}(y_t) +  \nu\, H_{AA}(y_t,\nu) \\ \nonumber
&=&  S_{NN}(y_t) +  \nu\,r_{AA}(y_t,\nu) \,H_{NN}(y_t),
\eea
where $S_{NN}$ ($\sim S_{pp}$) is the soft component and $H_{AA}$ is the A-A hard component (with reference $H_{NN}\sim H_{pp}$) 
~\cite{hardspec,ppprd}. Ratio $r_{AA} = H_{AA} / H_{NN}$ is an alternative ratio measure to nuclear modification factor $R_{AA}$. Centrality measure $\nu \equiv 2 n_{binary} / n_{participant}$ estimates the 
mean projectile-nucleon path length in A-A collisions. We are interested in the evolution of hard component $H_{AA}$ or ratio $r_{AA}$ with A-A centrality.
For the A-A two-component model the spectrum soft component remains by hypothesis unchanged and scales as the number of participant pairs $n_{part}/2$. For Glauber linear superposition of p-p (N-N) collisions (GLS reference) spectrum hard component $H_{AA} \rightarrow H_{NN}(y_t)$ would also remain unchanged modulo the factor $\nu$ relative to participant scaling. In real A-A collisions $H_{AA}(y_t,b)$ changes relative to GLS reference $H_{NN}(y_t)$, representing ``medium modification'' of parton fragmentation.

 Study of $H_{AA}(y_t,b)$ reveals evolution of fragmentation with centrality~\cite{hardspec}. pQCD can be used to calculate equivalent fragment distributions which can be compared directly with measured $H_{AA}$~\cite{fragevo}. FD calculations require a combination of {\em measured} fragmentation functions, a pQCD-predicted parton spectrum and a theoretical model of FF modification in A-A collisions.  Several sections below describe methods to calculate FDs and present interpretations of spectrum hard-component evolution based on theory-data comparisons.



 \section{Fragmentation functions}


We require a phenomenological representation of measured fragmentation functions. As noted, FFs are commonly plotted on momentum fraction $x_p = p_{fragment} / p_{jet}$ or $\xi_p = \ln(1/x_p)$.  Alternatively, one can define rapidity $y = \ln\{(E+p) / m_\pi\}$ with pion mass adopted for unidentified hadrons, where $p$ is the fragment {\em total} momentum appropriate for most FF data from $e^+$-$e^-$ collisions. 


 \begin{figure*}[h]
  \includegraphics[width=1.45in,height=1.65in]{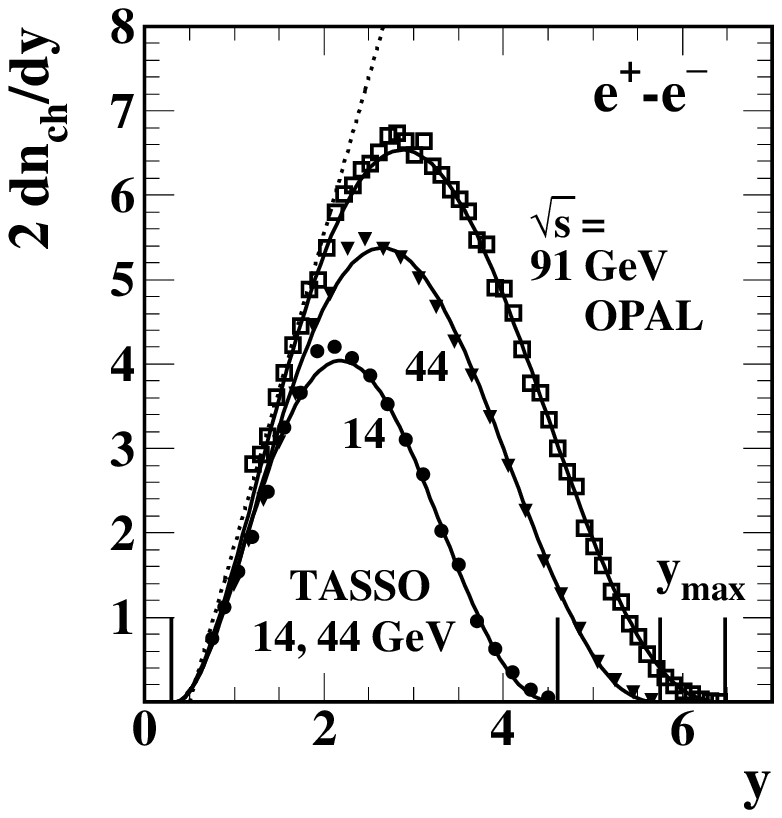}
  \includegraphics[width=1.45in,height=1.65in]{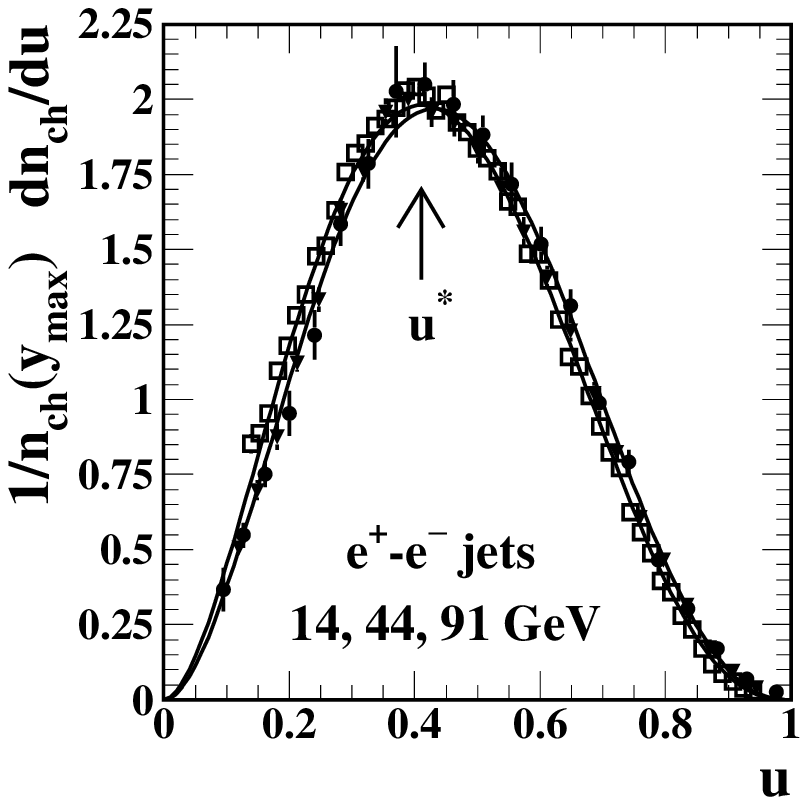}
  \includegraphics[width=1.45in,height=1.65in]{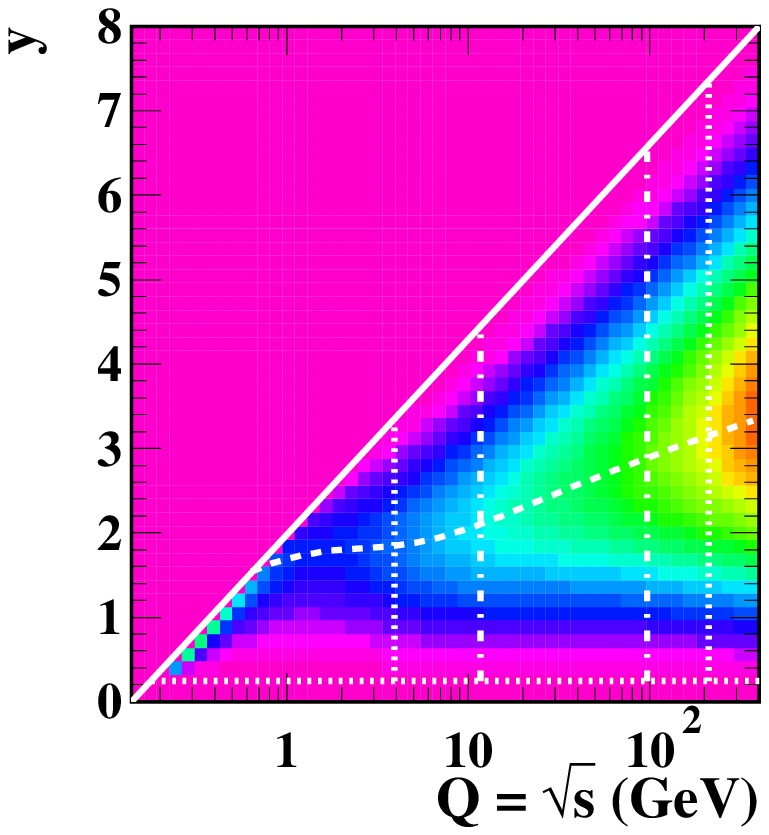}
  \includegraphics[width=1.45in,height=1.65in]{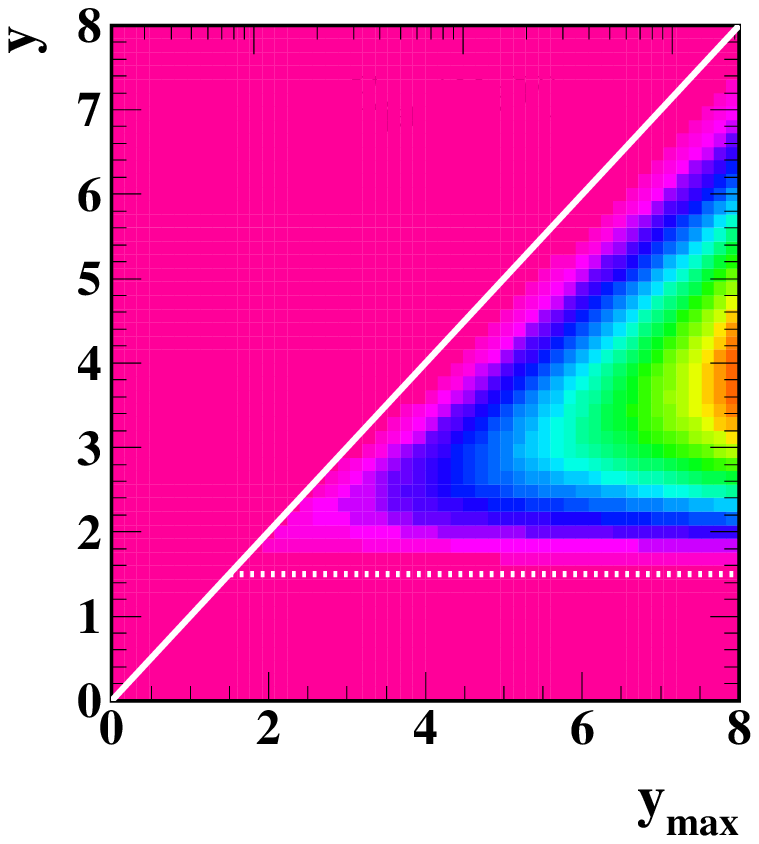}
\caption{\label{ffs}
First: Measured fragmentation functions (points) from $e^+$-$e^-$ collisions for three CM (dijet) energies~\cite{opal,tasso} plotted on rapidity $y$ with $\beta$-distribution parametrizations (solid curves).
Second: The same FF data and curves scaled to unit-normal distributions and plotted on normalized rapidity $u$.
Third: Parametrization of the ensemble of $e^+$-$e^-$ FFs (surface) over a large energy range.
Fourth: FF ensemble parameterization for p-\=p (p-p) collisions.
 } 
 \end{figure*}

Figure \ref{ffs} (first panel) shows FFs from $e^+$-$e^-$ collisions at three energies plotted as $D(x,Q^2) \leftrightarrow D(y,y_{max}) = 2dn_{ch}/dy$ on $y$~\cite{opal,tasso}. To good approximation the FFs are self-similar: both widths and amplitudes scale linearly with $y_{max} = \ln(Q/m_\pi) = \ln(2 E_{jet} / m_\pi)$~\cite{ffprd}. The jet fragment multiplicity $n_{ch,j}(y_{max})$ is  then approximately proportional to $y_{max}^2$. Deviations from that trend arise mainly from the running of $\alpha_s$. Because the FFs are nearly self-similar we can convert them to a universal form by renormalizing both the FF amplitude (to unit-normal) and the rapidity [to $u = (y - y_{min}) / (y_{max} - y_{min})$, where $y_{min} \sim 0.3$]. We then obtain the data in the second panel represented by a {\em beta} distribution (solid curves) to the error limits of the data: $D(y,y_{max}) \rightarrow 2n_{ch,j}(y_{max})\beta(u;p,q)$ (where $p,q$ are parton-energy-dependent parameters). Only the {\em best} FF data reveal any deviations from perfect $y_{max}$ ($Q^2$) scaling. The self similarity is the dominant aspect of DGLAP evolution. The simplicity is not apparent unless FFs are plotted on $y$. If the universal beta distribution is transformed back to individual jet energies we obtain the solid curves in the first panel and the surface in the third panel, which accurately describes all FFs above 3 GeV parton energy ($Q = 6$ GeV) and down to zero {\em hadron fragment} momentum~\cite{ffprd}. Each vertical slice of the surface plot is the FF for a particular parton energy. 

CDF p-\=p FF data (not shown) exhibit significant differences from $e^+$-$e^-$ FFs. The p-\=p FF ensemble is represented in the fourth panel as modified $e^+$-$e^-$ FFs. Part of the difference is due to an imposed cone radius which should exclude some low-momentum fragments. However, the p-\=p data suggest a real reduction relative to $e^+$-$e^-$ FFs. The evolution with energy scale of p-\=p FFs is also anomalous: there is a saturation of the FF amplitudes at larger jet energies compared to LEP FFs. FF universality may not be a valid assumption given the energy trend of those data~\cite{fragevo}.

 \section{Parton spectrum}

\begin{figure*}[h]
  \includegraphics[width=1.46in,height=1.65in]{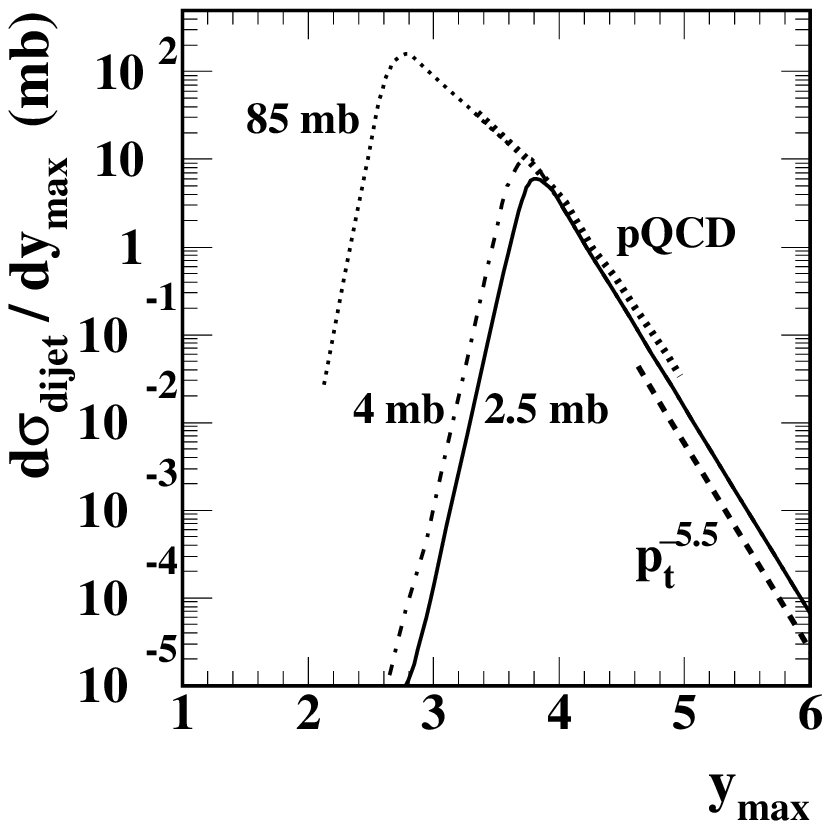}
  \includegraphics[width=1.46in,height=1.65in]{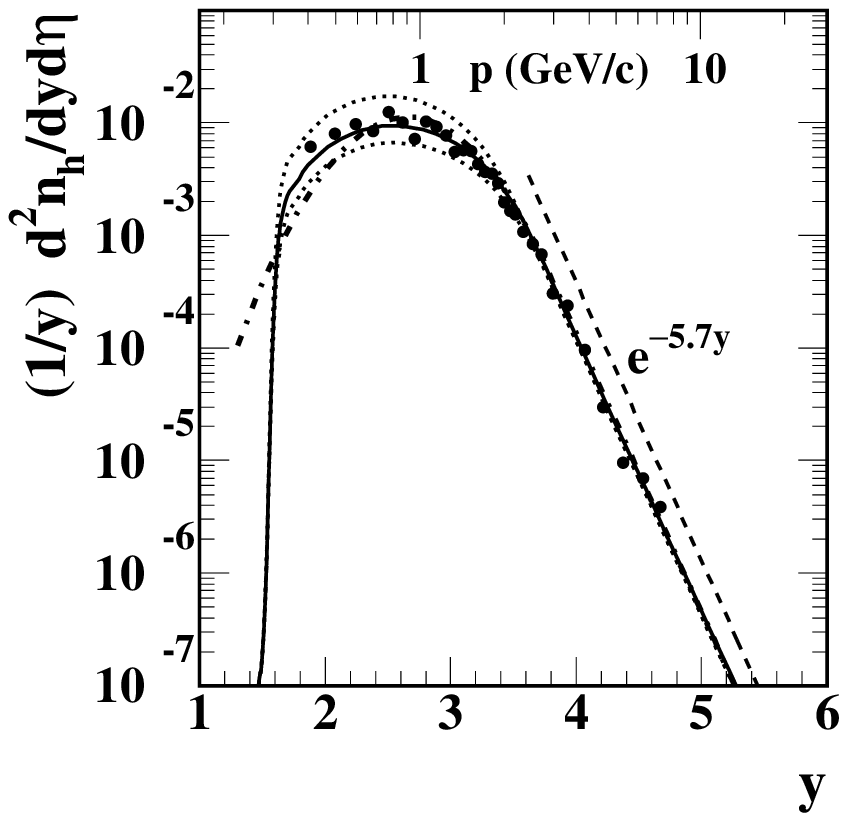}
 \includegraphics[width=1.46in,height=1.65in]{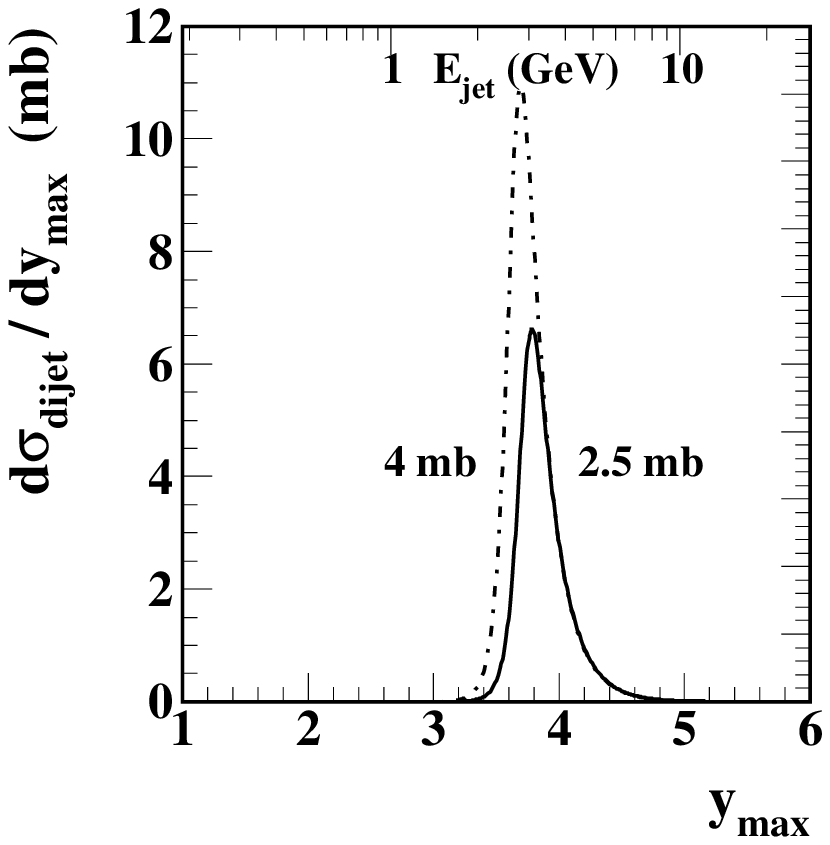}
   \includegraphics[width=1.46in,height=1.65in]{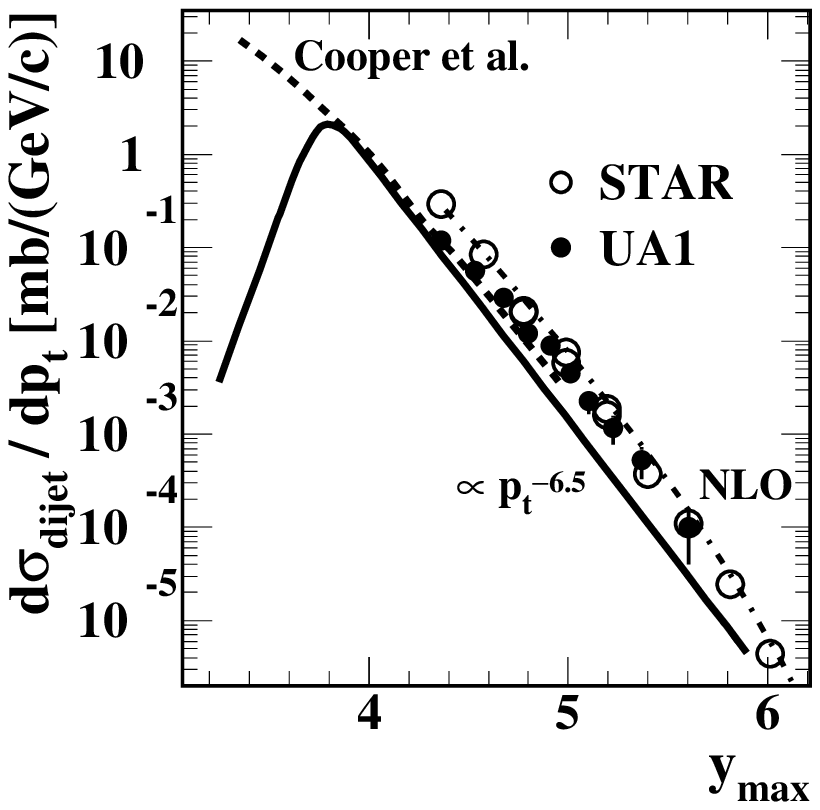}
\caption{\label{pspec}
 First: Parton spectra inferred from this analysis for p-p collisions (solid curve) and central Au-Au collisions (dash-dotted curve) compared to an ab-initio pQCD theory result (bold dotted curve~\cite{cooper}).
Second: Fragment distribution (solid curve) compared to p-p hard-component data (points). Dotted curves correspond to $\pm$10\% change in parton spectrum cutoff energy about 3 GeV. 
Third: Curves in the first panel plotted on a linear scale to illustrate that almost all partons (gluons) appear near 3 GeV.
Fourth: Comparison of parton spectrum inferred in Ref.~\cite{fragevo} (solid curve) with ab initio pQCD calculations (bold dashed and dot-dashed curves) and event-wise reconstructed jet spectra (data points)~\cite{ua1,starpp}.
} 

 \end{figure*}

Next we require a pQCD parton spectrum. In Fig.~\ref{pspec} (first panel) the solid curve is a power-law spectrum with energy cutoff inferred by working backward from 200 GeV p-p and A-A spectrum hard-component data according to procedures described below. The bold dotted curve is an ab initio pQCD calculation~\cite{cooper}. The two spectra agree quantitatively near 3 GeV ($y_{max} \sim 3.8$) where almost all scattered partons appear. The good agreement, arising from two independent approaches to parton spectrum determination, is significant. Parton spectrum details at larger parton energies are less important for {\em minimum-bias} spectrum and correlation analysis. 

Given the power-law approximation to the parton spectrum it is not apparent from pQCD where the effective spectrum cutoff should be. Figure~\ref{pspec} (second panel) shows calculated FDs from the procedure described below with spectrum cutoffs at $3 \pm 0.3$ GeV (solid and dotted curves) compared to the p-p spectrum hard-component data introduced above. The comparison establishes an empirical 3 GeV spectrum cutoff. Figure~\ref{pspec} (third panel) shows the power-law spectrum (with cutoff) on a linear scale, demonstrating that almost all scattered partons appear at the cutoff energy.

Saturation-scale (SS) arguments support a cutoff at 1 GeV (light dotted curve in the first panel)~\cite{cooper}. Given the approximate power-law dependence the difference in the total parton yield for the two cutoffs is a factor 30-50 in the initial parton (mainly gluon) density. The 1 GeV SS cutoff is based on an argument derived from initial-state parton densities in nucleons. Considered as a quantum-mechanical process parton scattering and fragmentation to charged hadrons depends not only on the initial-state parton density but also on the final-state hadron density of states. If there is no final state for a given parton scattering the transition is not allowed. The effective cutoff should then depend on the available density of hadronic final states at a given parton energy scale. 

Figure~\ref{pspec} (fourth panel) shows data from a UA1 analysis of energy clusters in EM calorimeter data (solid dots) leading to inference of ``minijets,'' with spectrum extending down to 5 GeV (later amended to 3-4 GeV after background subtraction)~\cite{ua1}. More recent data from the STAR collaboration (open circles) extending down to 5 GeV (4 GeV without background contribution) are also shown~\cite{starpp}.  The parton spectrum inferred phenomenologically from RHIC p-p data (solid curve) includes a cutoff near 3 GeV which is consistent with the UA1 observations of 1985 and with the STAR reconstructed-jet spectrum. The undershoot of the solid curve at larger parton energy may be due to oversimplified modeling of p-\=p FFs. If the p-\=p FF saturation mentioned above is included the inferred parton spectrum should be even closer to pQCD theory and jet spectra. The dash-dotted curve is a spectrum from Ref.~\cite{nlo}.



 \section{$\bf pQCD$ folding integral and fragment distributions}




The pQCD folding (convolution) integral used to calculate fragment distributions is
\bea \label{fold}
\frac{d^2n_{h}}{dy\, d\eta}   &\approx&    \frac{\epsilon(\Delta \eta)/2}{ \sigma_{_{\tiny NSD}}\, \Delta \eta_{4\pi}}  \int_0^\infty   dy_{max}\, D(y,y_{max}) \frac{d\sigma_{dijet}}{dy_{max}},
\eea
where $D(y,y_{max})$ is the FF ensemble  for a specific collision system ($e^+$-$e^-$, p-p, A-A, in-medium or in-vacuum), and $d\sigma_{dijet}/dy_{max}$ is the parton spectrum~\cite{fragevo}. The perturbative object is the parton spectrum; the nonperturbative object is the measured FF ensemble. The folding integral then produces a prediction for the observed hadron spectrum hard component. ${d^2n_{h}}/{dy\, d\eta}$ is the predicted hadron FD from parton pairs  scattered into  angle acceptance $\Delta \eta$. 
Efficiency factor $\epsilon \in [1,2]$ 
includes the possibility that the second jet of a dijet also falls within $\Delta \eta$. 
$\Delta \eta_{4\pi} \sim 5$ is the effective $4\pi$ $\eta$ interval for scattered partons~\cite{ua1}, and $\sigma_{NSD}\sim 36$ mb is the cross section for NSD p-p collisions, both for $\sqrt{s_{NN}} = 200$ GeV.


 \begin{figure*}[h]
  \includegraphics[width=1.46in,height=1.65in]{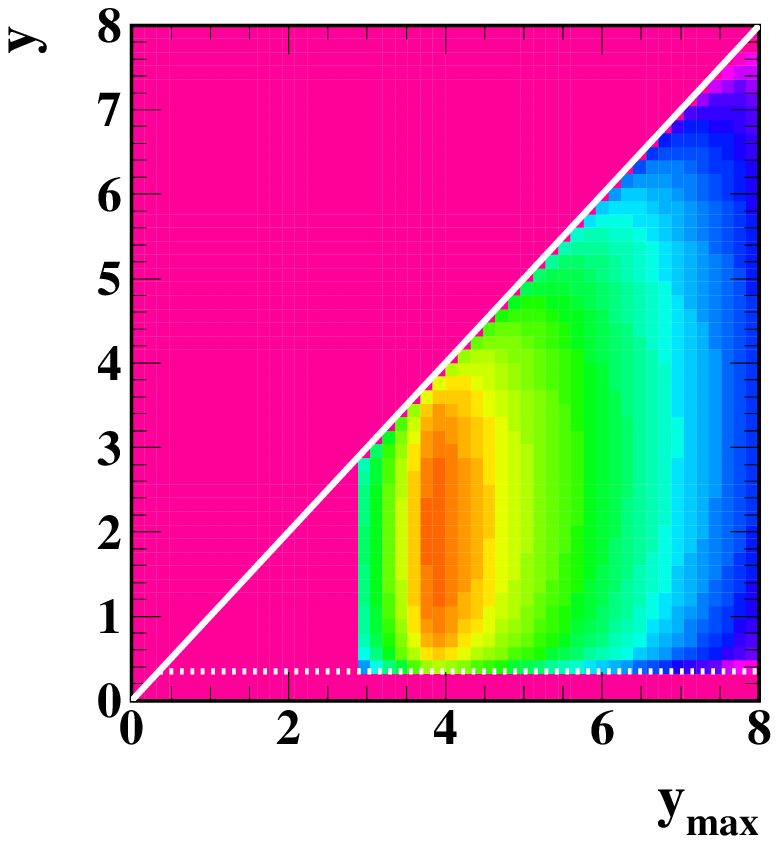}
  \includegraphics[width=1.46in,height=1.65in]{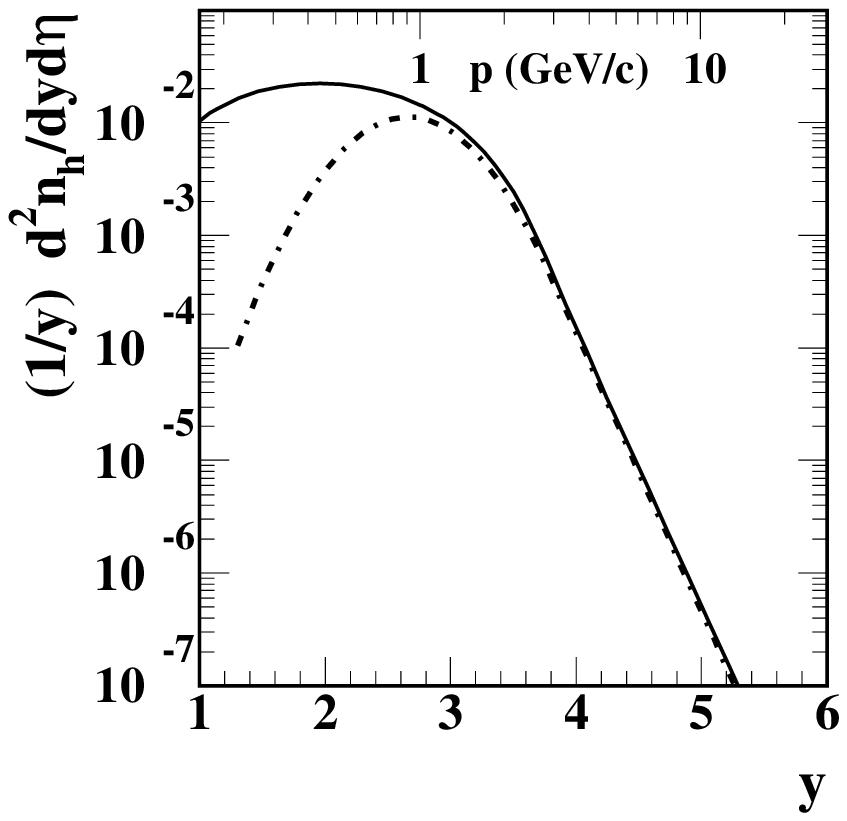}
   \includegraphics[width=1.46in,height=1.65in]{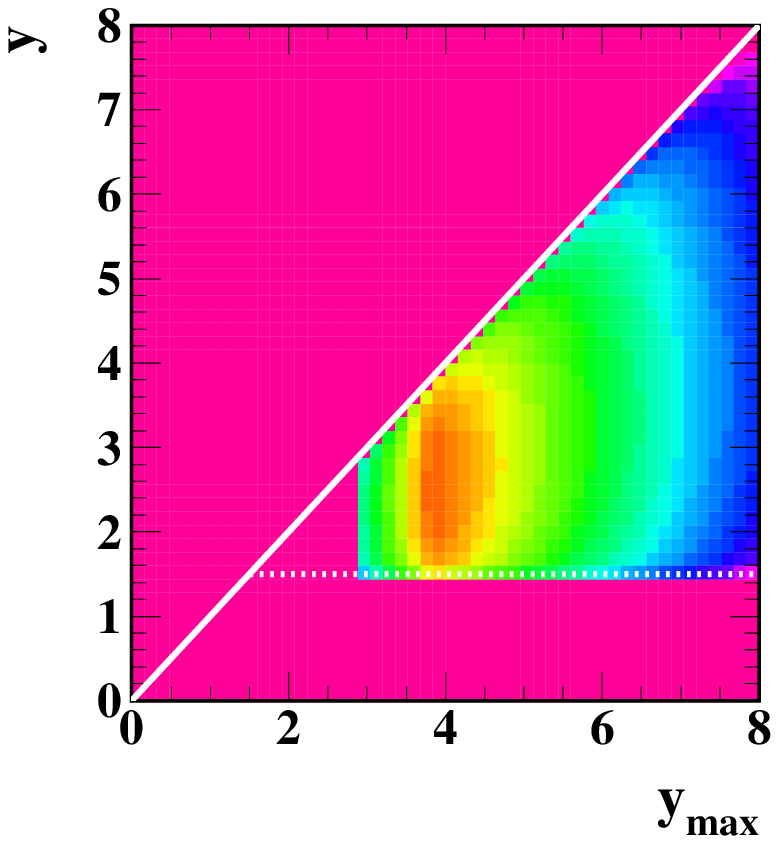}
 \includegraphics[width=1.46in,height=1.65in]{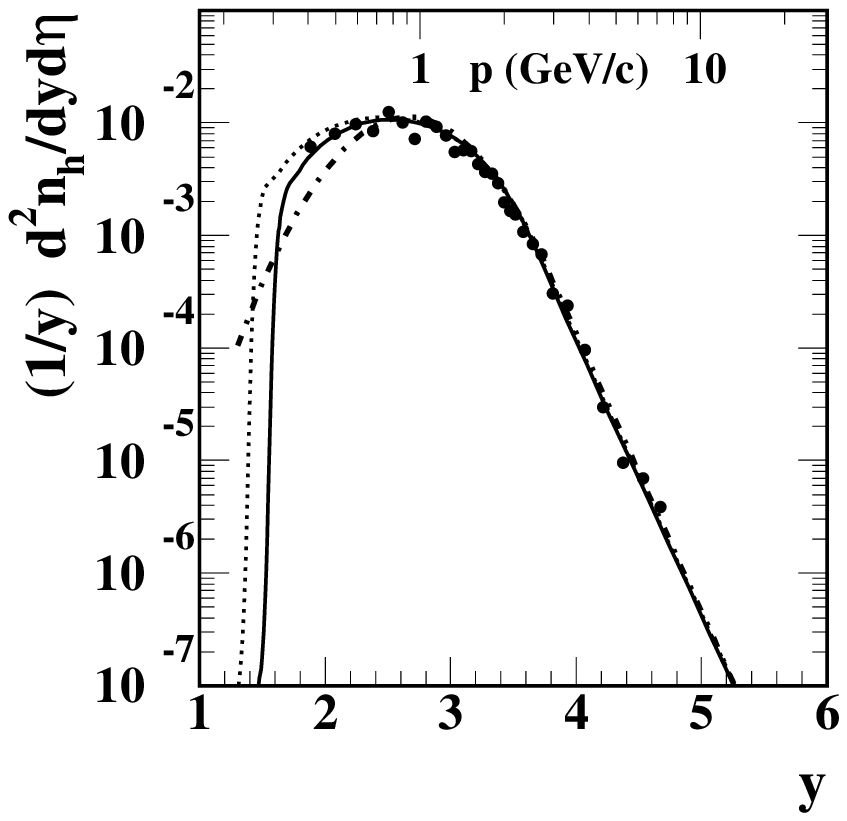}
\caption{\label{folding}
First: pQCD folding-integral argument for $e^+$-$e^-$ FFs.
Second: $e^+$-$e^-$ FD (solid curve) and p-p hard-component reference from Ref.~\cite{ppprd} (dash-dotted curve).
Third: Folding-integral argument for p-\=p FFs.
Fourth: p-p FD (solid curve), p-p hard-component data (solid dots) and reference (dash-dotted curve).
 } 
 \end{figure*}

Figure~\ref{folding} (first panel) shows the folding-integral argument $D(y,y_{max}) d\sigma_{dijet}/dy_{max}$ for  $e^+$-$e^-$ FFs and 3 GeV parton spectrum cutoff. The fragment distribution is then the projection of the 2D histogram onto fragment rapidity $y$. Figure~\ref{folding} (third panel) shows the argument for FFs from p-\=p collisions with cutoff on fragment rapidity higher than for  $e^+$-$e^-$ collisions. 
Figure~\ref{folding} (second panel) shows the FD projection (solid curve) compared to a model of the measured spectrum hard component in NSD p-p collisions (dash-dotted curve). The FD from $e^+$-$e^-$ FFs is adopted below as the reference for all measured hard-component distributions. In general, fragment distributions (FDs) from theory are compared to spectrum hard components (HCs) from data. Figure~\ref{folding} (fourth panel) shows the FD (solid curve) for p-\=p collisions compared to the HC (points) from NSD p-p collisions~\cite{ffprd}. That comparison established the 3 GeV parton spectrum cutoff~\cite{fragevo}.

According  to these calculations the most significant variations in fragmentation and the largest fragment yields appear below 2 GeV/c, which reveals a {\em fundamental logical problem} in the conventional RHIC approach to data analysis and interpretation.  The spectrum interval below 2 GeV/c ($y_t = 3.3$), described as the ``soft'' region of the hadron spectrum, is conventionally assigned to hydro models. The interval above 6 GeV/c is conventionally assigned to ``hard processes'' (parton scattering and fragmentation) described by pQCD. The source of the logical difficulty lies in confusing theoretical limitations on pQCD descriptions of fragmentation functions with the theoretical ability to describe fragment distributions in terms of {\em measured} FFs. It is the parton spectrum which must be described perturbatively, not fragmentation functions. By imposing an unjustified constraint on pQCD descriptions of fragmentation the great majority of hadron fragments is surrendered to hydro interpretations.





 \section{Parton ``energy loss'' and medium-modified fragment distributions}

\begin{figure*}[h]
\includegraphics[width=1.46in,height=1.65in]{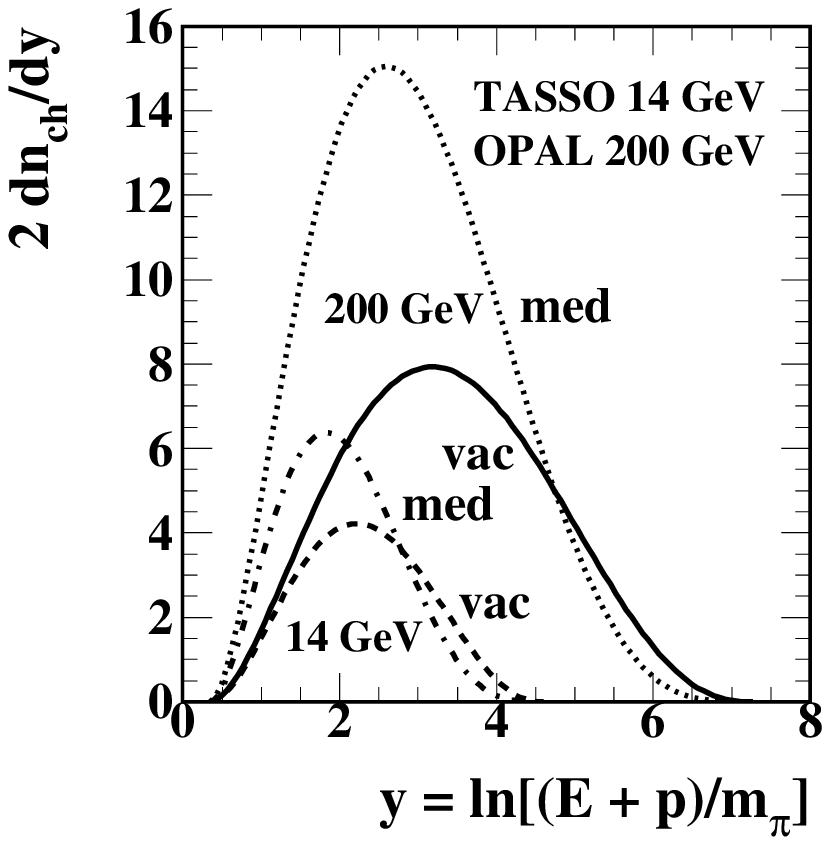} 
\includegraphics[width=1.46in,height=1.67in]{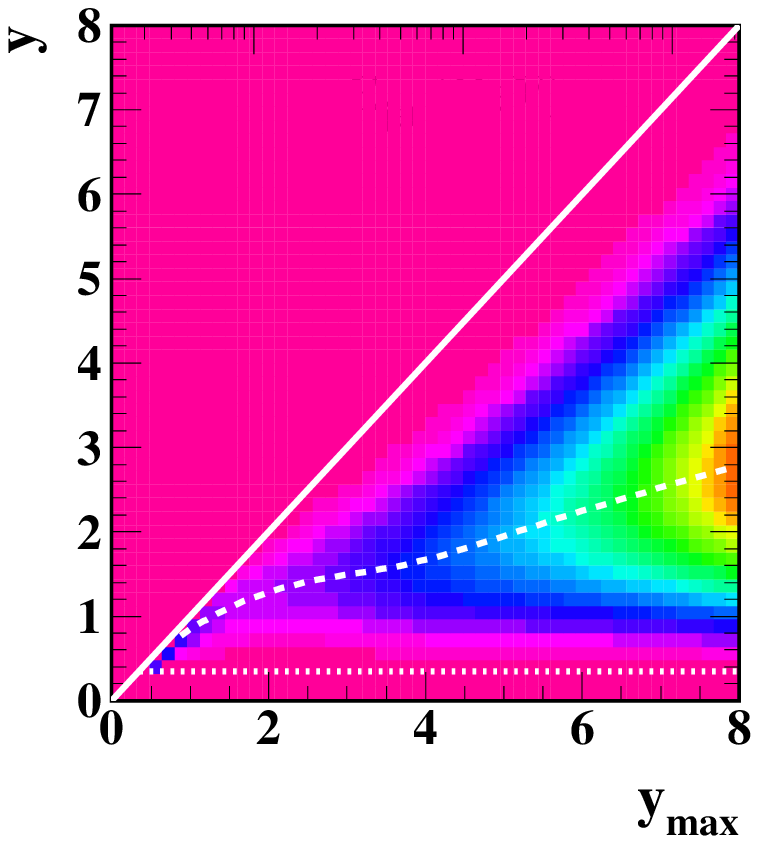}
\includegraphics[width=1.46in,height=1.65in]{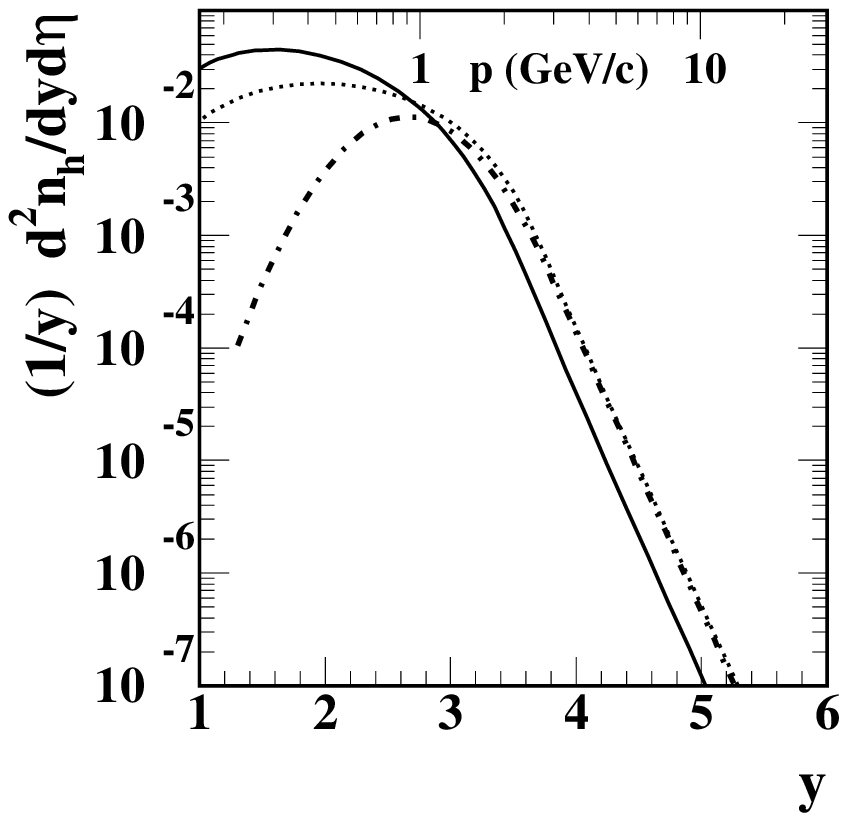}
\includegraphics[width=1.46in,height=1.65in]{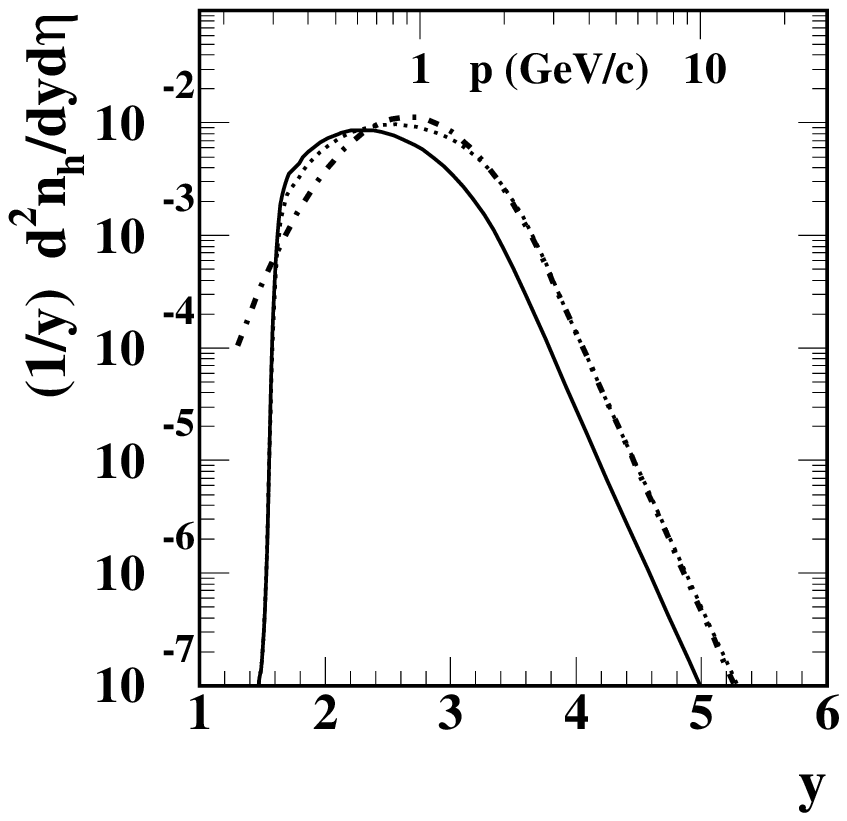}
\caption{\label{ffmods}
First: $e^+$-$e^-$ FFs for two energies unmodified~\cite{ffprd} (solid and dashed curves) and modified to emulate parton ``energy loss''~\cite{borg} (dash-dotted and dotted curves).
Second: Modified $e^+$-$e^-$ FF ensemble.
Third: Medium-modified FD from $e^+$-$e^-$ FFs (solid curve) compared to in-vacuum FD (dotted curve).
Fourth: Medium-modified FD from p-\=p FFs (solid curve) compared to in-vacuum FD (dotted curve).
}
\end{figure*}

We next require a model for medium modification of fragmentation functions. Figure~\ref{ffmods} (first panel) shows ``medium modified'' fragmentation functions (Borghini-Wiedemann, BW) achieved by altering certain splitting functions in the parton cascade~\cite{borg}. The solid and dashed curves (vac) are parametrizations of $e^+$-$e^-$ FFs based on the beta distribution which describe FF data within their uncertainties down to zero fragment momentum~\cite{ffprd}. The dotted and dash-dotted curves (med) are FFs modified to match the BW prescription simply by changing  the parameter $q$ in the beta distribution which, by construction, {\em conserves the parton energy}. There is suppression of larger-momentum fragments and consequent enhancement of smaller-momentum fragments~\cite{fragevo}. 

The modified $e^+$-$e^-$ FF ensemble is shown in the second panel. The locus of modes (white dashed curve) is shifted to smaller fragment rapidities $y$ (compare with Fig.~\ref{ffs}, third panel). There is a similar result for the p-p (p-\=p) FF ensemble. By inserting modified FFs into the pQCD folding integral Eq.~(\ref{fold})  we obtain modified fragment distributions (solid curves) compared to unmodified FDs (dotted curves) in the third and fourth panels~\cite{fragevo}.
At larger $p_t$ there is ``jet suppression'' (parameter $q$ is adjusted to match spectrum data there).  At smaller $p_t$ there is corresponding ``jet enhancement,'' a new aspect of the fragmentation problem. The low-$p_t$ enhancement is large for $e^+$-$e^-$ FFs but negligible for p-\=p FFs. That difference becomes important in A-A collisions.

 \section{Evolution of fragmentation with centrality} \label{evolution}

\begin{figure*}[h]
  \includegraphics[width=1.46in,height=1.65in]{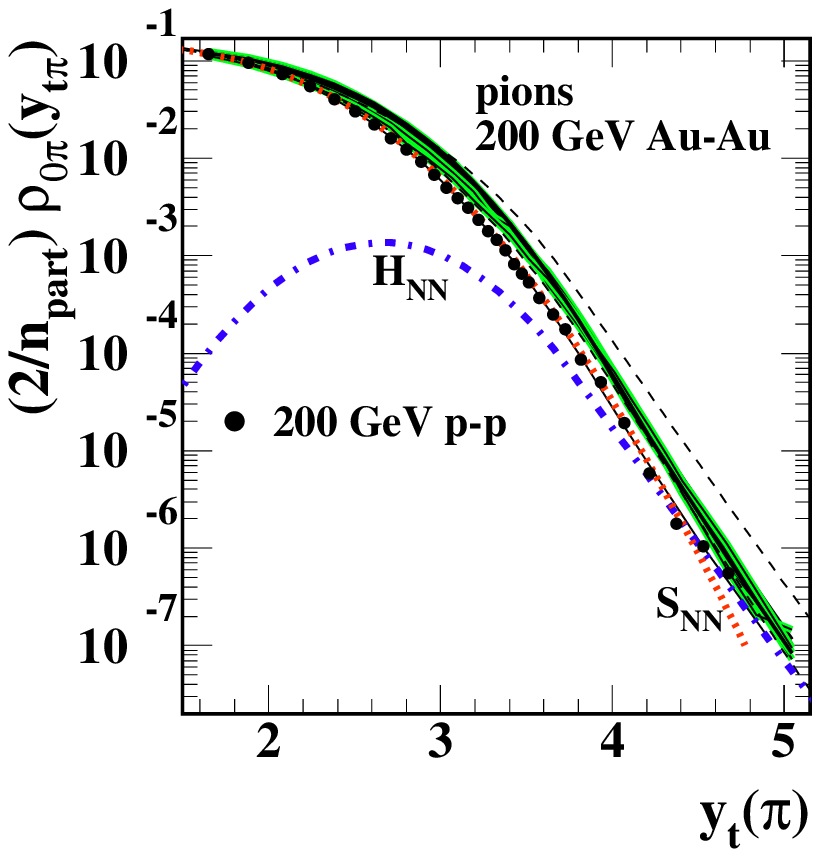}
  \includegraphics[width=1.46in,height=1.65in]{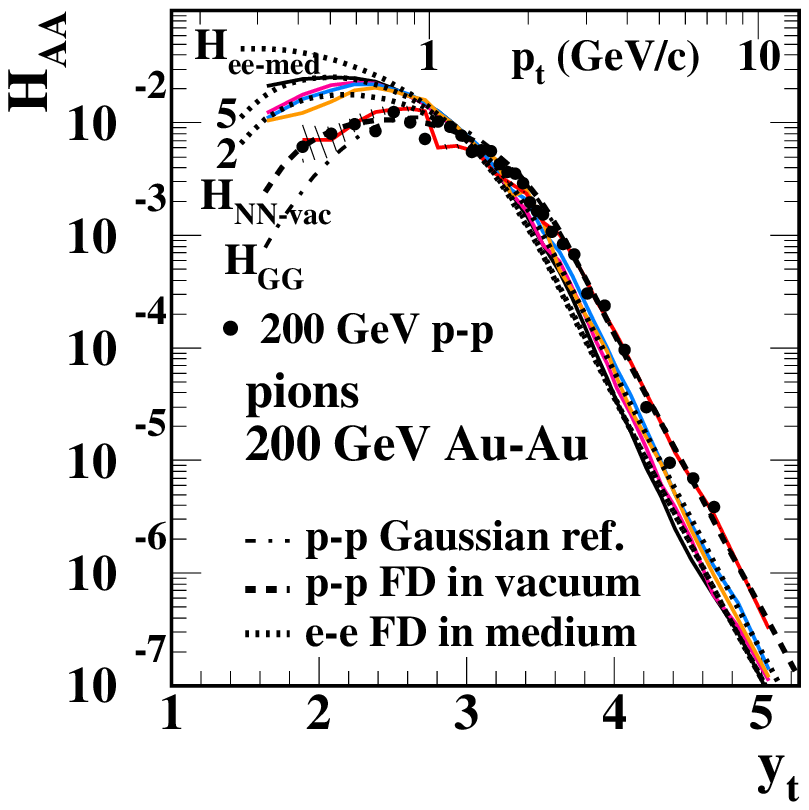}
   \includegraphics[width=2.92in,height=1.65in]{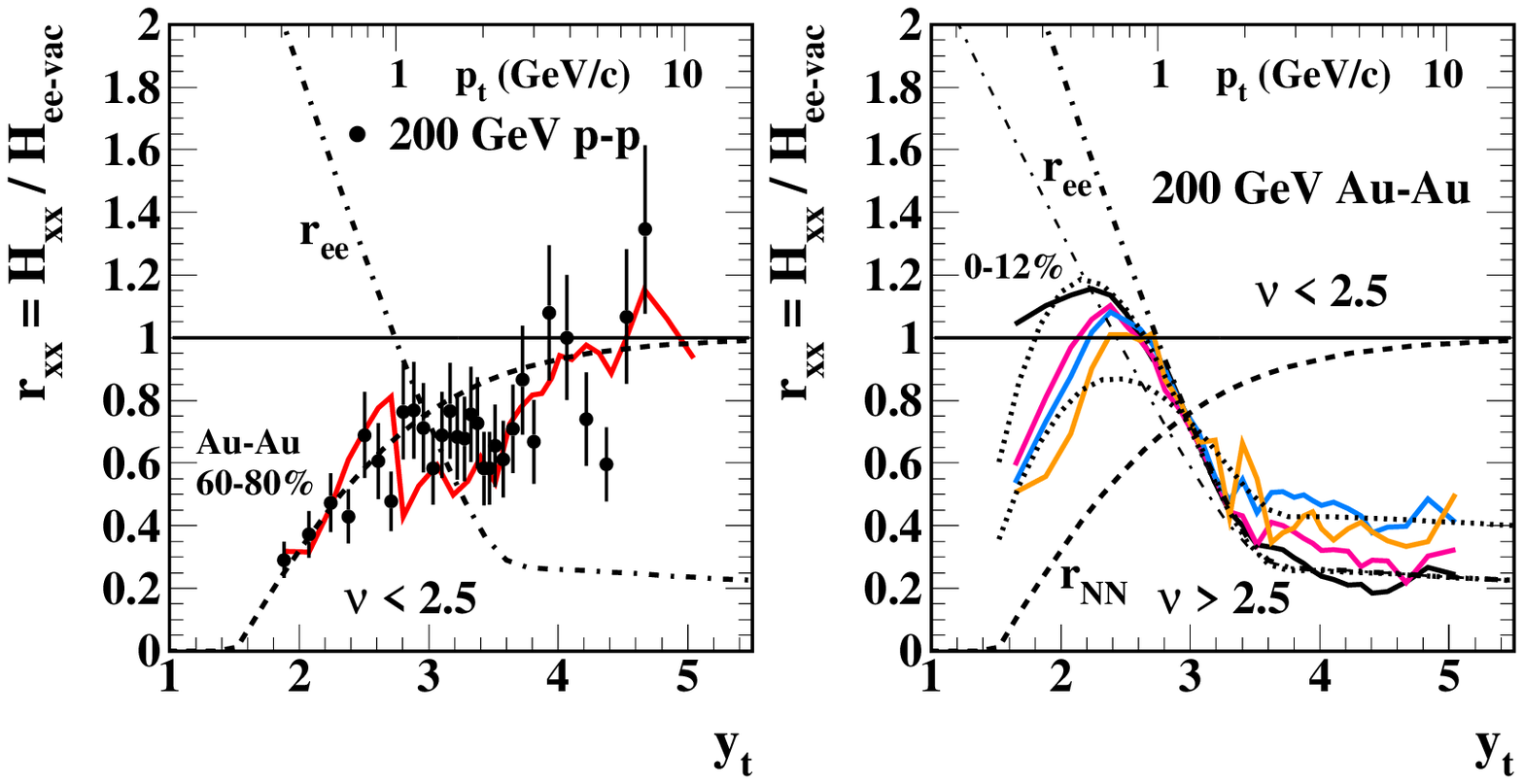}
\caption{\label{fragevo}
First: Pion spectra (dark solid curves) for five centralities of 200 GeV Au-Au collisions.
 Second: Hard-component centrality evolution in Au-Au collisions~\cite{hardspec}. Large enhancements  at smaller $y_t$ accompany suppression at larger $y_t$.
Third: Hard-component (HC) ratios relative to an ee-vacuum reference for Au-Au collisions below the sharp transition.
Fourth: HC ratios above the sharp transition revealing major changes in HC structure.
} 
 \end{figure*}

We can now make a direct comparison between calculated pQCD FDs  and measured spectrum HCs.
Figure~\ref{fragevo} (first panel) shows measured pion spectra from 200 GeV Au-Au for five centralities plotted on pion rapidity (dark solid curves)~\cite{hardspec}. $S_{NN}$ is the common soft component inferred as the limiting spectrum for centrality measure $\nu \rightarrow 0$. The solid dots are the hadron spectrum from NSD p-p collisions~\cite{ppprd}. The dash-dotted curve is hard component $H_{NN}$ inferred from the $n_{ch}$ dependence of the p-p spectrum. 


Figure~\ref{fragevo} (second panel) shows $H_{AA}$ extracted from the Au-Au spectra (bold solid curves) according to Eq.~(\ref{aa2comp}): we subtract the same soft component from spectra for five centralities and divide by $\nu$. Although the soft component certainly dominates spectra below 1 GeV/c the systematic uncertainty in $H_{AA}$ is manageable at least down to 0.5 GeV/c ($y_t = 2$). The fixed soft component plus the inferred hard components describe the original spectrum data exactly. Relative to GLS reference $H_{NN}$ (bold dashed curve) there is suppression at larger $p_t$ and enhancement at smaller $p_t$ in more-central Au-Au collisions. Detailed systematic study of HC evolution with Au-Au centrality reveals that with increasing centrality i) p-\=p FFs transition to $e^+$-$e^-$ FFs, ii) FFs become ``medium modified'' and iii) there is a 50\% increase in the dijet cross section due to a 10\% reduction in the effective parton spectrum cutoff energy (3 GeV $\rightarrow$ 2.7 GeV). The bold dotted curves in the second panel show pQCD FDs calculated according to those systematic trends~\cite{fragevo}.

To obtain a more differential view of HC evolution we define ratios $r_{xx} = H_{xx} / H_{ref}$, where $xx$ denotes the collision system and $H_{ref}$ is {\em not} $H_{NN}$ inferred from p-p collisions. Instead, $H_{ref} \rightarrow H_{ee-vac}$ is defined by an FD constructed according to Eq.~(\ref{fold}) using the parton spectrum inferred from p-p collisions~\cite{fragevo} and in-vacuum FFs from $e^+$-$e^-$ collisions~\cite{ffprd}. 
Figure~\ref{fragevo} (third panel) shows $r_{xx}$ data for the HC from p-p collisions (solid dots) and from more-peripheral Au-Au collisions (bold solid curve). Reference curve $r_{NN}$ (dashed) is obtained from the FD for N-N collisions using p-\=p FFs ($\sim H_{NN}$).  Dash-dotted reference curve $r_{ee}$ is obtained from the FD for in-medium modified $e^+$-$e^-$ FFs corresponding to central Au-Au collisions (parton spectrum cutoff reduced by 10\%). 


Figure~\ref{fragevo} (fourth panel) shows $r_{xx}$ data (bold solid curves) for HCs from more-central Au-Au collisions. We observe a {\em sharp transition} in HC evolution at $\nu = 2.5$, with qualitatively different behavior below and above the transition. Below the transition we observe strong suppression of  the HC at smaller $p_t$ relative to what is expected for in-vacuum $e^+$-$e^-$ FFs (third panel).
Above the transition, in more-central Au-Au collisions, we observe a strong enhancement at smaller $p_t$ complementing the suppression at larger $p_t$ observed with conventional ratio $R_{AA}$. The HC centrality dependence at 0.5 GeV/c corresponds in detail to that at 10 GeV/c. The number of particles at smaller $p_t$ resulting from parton fragmentation is much greater than that at larger $p_t$ (consistent with approximate parton energy conservation). The copious low-$p_t$ hadron fragments should be accommodated in any theoretical description of A-A collisions. 

 \section{Fragment yields from jet angular correlations}



We can infer fragment yields from $p_t$-integrated jet angular correlations via factorization of the measured jet-correlated pair density to obtain the per-jet fragment multiplicity. We integrate Eq.~(\ref{fold}) over fragment rapidity $y$ on both sides to obtain $dn_h/ d\eta$, the per-participant-pair fragment density on $\eta$, in terms of {\em jet frequency} $f(b)$ and mean {\em jet fragment multiplicity} $n_{ch,j}(b)$
\bea \label{fold2}
\int dy\, \frac{d^2n_{h}}{dy\,d\eta}   &=&    \left\{\frac{\epsilon(\Delta \eta)\, \sigma_{dijet}(b)}{\sigma_{NSD}\, \Delta \eta_{4\pi}} \right\} \left\{ \frac{1}{\sigma_{dijet}(b)} \int_0^\infty   dy_{max}\, n_{ch,j}(y_{max},b)\, \frac{d\sigma_{dijet}}{dy_{max}} \right\} \\ \nonumber
 &\equiv&   f(b)\,n_{ch,j}(b).
\eea
$n_{ch,j}(b)$, averaged over the minimum-bias parton spectrum, is effectively the mean fragment multiplicity for partons near the parton spectrum cutoff ($\sim$3 GeV). We can infer $n_{ch,j}(b)$ from jet angular correlations for A-A centrality $b$.
Jet frequency $f(b) = (1/n_{bin})\,dn_j(b)/d\eta$ is the number of jets per unit $\eta$ per NSD N-N collision estimated from pQCD. The argument of $\sigma_{dijet}(b)$ admits the possibility that the N-N dijet cross section may depend on A-A centrality~\cite{fragevo}.

 \subsection{Jet angular correlations}

\begin{figure*}[h]
  \includegraphics[width=1.46in,height=1.65in]{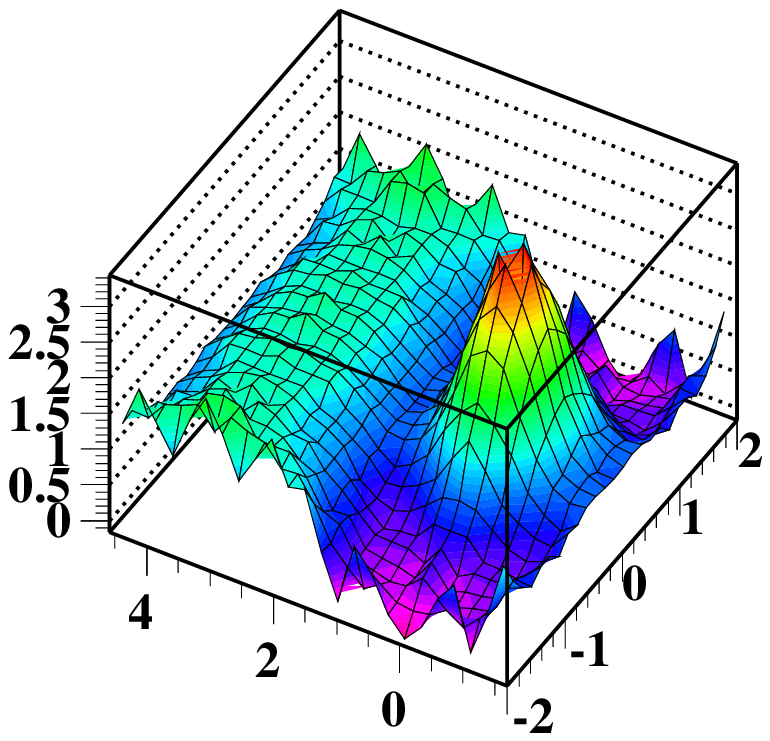}
  \includegraphics[width=1.46in,height=1.65in]{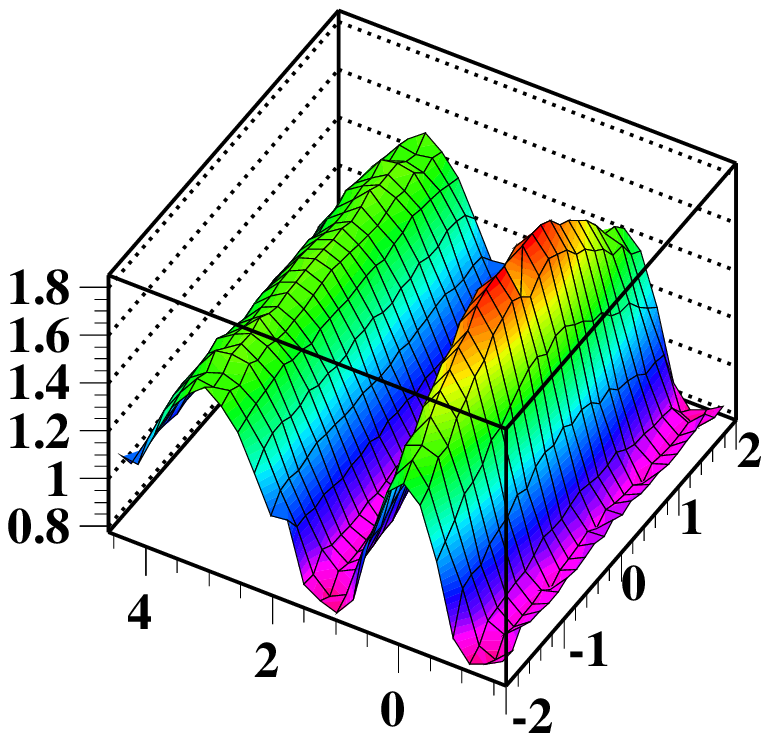}
   \includegraphics[width=1.46in,height=1.65in]{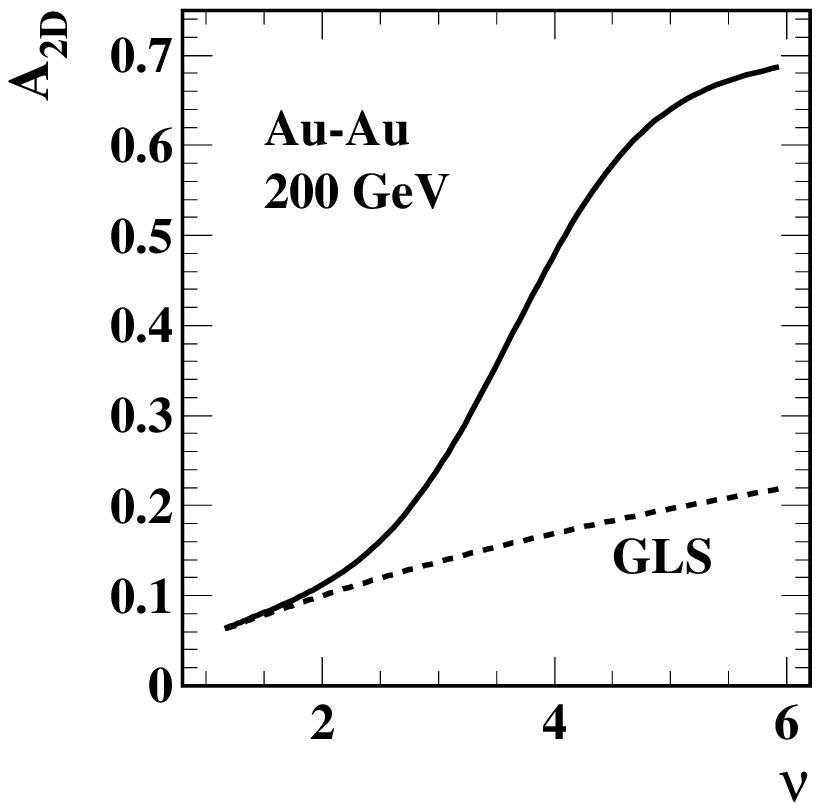}
   \includegraphics[width=1.46in,height=1.65in]{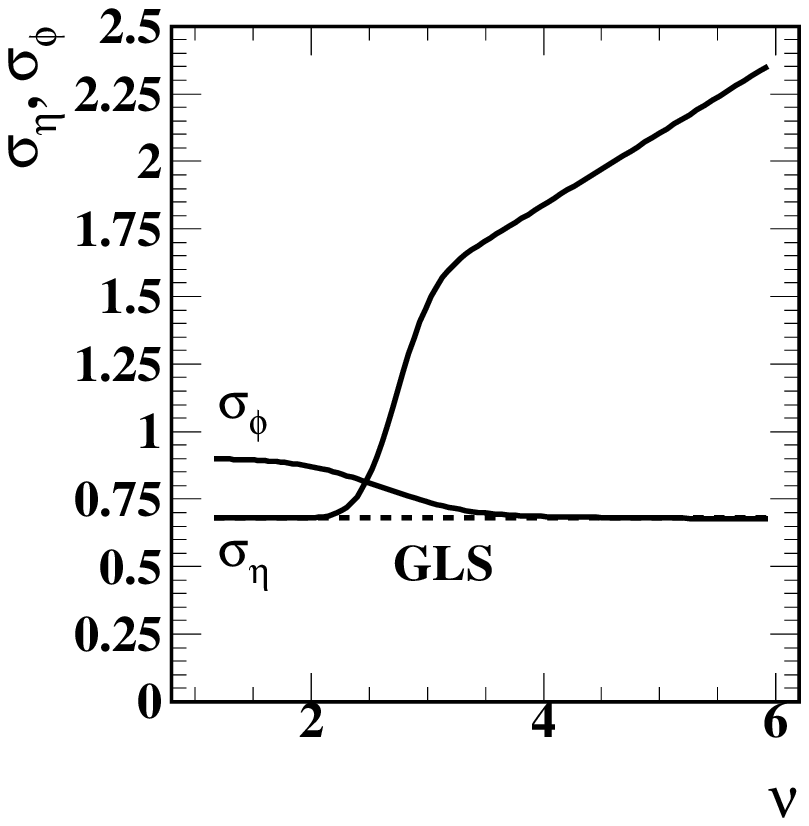}
\caption{\label{sspeak}
  First: 2D angular correlations 
from more-peripheral ($\nu \sim 1.4$) 200 GeV Au-Au collisions.
Second: 2D angular correlations from more-central ($\nu \sim 4.8$) 200 GeV Au-Au collisions.
Third: Same-side 2D peak amplitude (solid curve) compared to GLS reference (A-A transparency, dashed curve).
Fourth: Same-side 2D peak widths (solid curves) compared to GLS references (dashed curves).
} 
 \end{figure*}

2D angular autocorrelations on {difference variables} $\eta_\Delta = \eta_1 - \eta_2$ and $\phi_\Delta = \phi_1 - \phi_2$ evaluated near mid-rapidity retain all angular correlation information~\cite{inverse}. 2D correlations can be constructed for the $p_t$-integral minimum-bias case or with specific $p_t$ cuts on one or both particles in a pair. 2D angular correlation histograms are formed for p-p collisions~\cite{porter1,porter2} and several (typically 11) centrality classes of A-A collisions~\cite{daugherity}.
Figure~\ref{sspeak} (left panels) shows 2D histograms for peripheral ($\nu = 1.4$, $\sim$ p-p) and mid-central ($\nu \sim 4.8$) 200 GeV Au-Au collisions.
The correlation structure has three main features: a same-side (SS, $\phi_\Delta < \pi/2$) 2D peak at the origin, an away-side (AS, $\phi_\Delta > \pi/2$) ridge approximately uniform on $\eta_\Delta$ and described by dipole $\cos(\phi_\Delta - \pi)$ in more-central A-A collisions, and {\em non-jet} azimuth quadrupole $\cos(2\phi_\Delta)$. 
Angular correlations in Ref.~\cite{daugherity} are reported in the {\em per-particle} form  $\Delta \rho/\sqrt{\rho_{ref}}$~\cite{inverse}.
The SS jet peak is modeled by a 2D Gaussian
\bea \label{estruct}
\frac{\Delta \rho_{SS}}{{\sqrt{\rho_{ref}}}} &\equiv& \rho_0(b)\, j^2(\eta_\Delta,\phi_\Delta,b) 
 =  A_{2D}\, \exp\left(  -\eta^2_\Delta / 2\sigma^2_{\eta} \right)\, \exp\left(  -\phi^2_\Delta / 2\sigma^2_{\phi}\right),
\eea
where $j^2$ is a sibling/mixed pair ratio and $\rho_0(b)$ is the single-particle 2D angular density at mid-rapidity.

In Fig.~\ref{sspeak} (left panels) the fitted non-jet quadrupole has been subtracted. The SS peak can be interpreted as {\em intra}jet correlations and should include all hadron fragment pairs from jets that survive partonic and hadronic rescattering.
The AS ridge can be interpreted as {\em inter}jet correlations from back-to-back scattered partons. 
The most probable $p_t$ for minimum-bias jet-correlated pairs is 1 GeV/c in p-p collisions, consistent with the mode of the spectrum hard component.

SS 2D peak properties vary strongly with centrality. Figure~\ref{sspeak} (right panels) shows SS peak parameter variations with centrality parameter $\nu$~\cite{daugherity}. We observe a sharp transition in SS peak properties at $\nu \sim 2.5$ in 200 GeV Au-Au collisions which corresponds to the spectrum hard-component systematics noted in Sec.~\ref{evolution}. Below the transition the peak properties follow Glauber linear superposition of p-p structure as expected for transparent A-A collisions.  Above the transition the SS peak amplitude increases rapidly relative to GLS, there is strong elongation on $\eta$ and slight narrowing on $\phi$. The narrowing on azimuth is inconsistent with parton energy-loss models based on multiple scattering. It is notable that even in more-central Au-Au collisions (e.g. Fig.~\ref{sspeak}, second panel) the SS peak is well-described by a 2D Gaussian, and the AS peak on azimuth is an undistorted dipole~\cite{tzyam}.
SS peak systematics appear to correspond to jets, but what do those correlation trends imply for single-particle yields in the final state?  To answer that question we convert (factorize) two-particle jet correlations to obtain the equivalent in single-particle hadron fragment yields and spectrum hard components.




The SS peak volume is by hypothesis the number of jets in the angular acceptance times the number of fragment pairs per jet, which allows us to factorize the SS jet peak. To convert from jet angular correlations to parton fragment yields and spectra requires four steps: i) angle-average the SS 2D peak on $(\eta_\Delta,\phi_\Delta)$ over the 4D angular acceptance on $(\eta_1,\eta_2,\phi_1,\phi_2)$ to obtain mean pair ratio $j^2(b)$, ii) estimate the mean pQCD jet number per event $n_j(b)$ within the $\eta$ acceptance, iii) calculate the mean per-jet fragment multiplicity $n_{ch,j}(b)$, iv) combine those elements to infer jet fragment yields/spectra as HC 2D densities on $(\eta,\phi)$. The result of step i), the average of the SS 2D peak described by Eq.~(\ref{estruct}) over the angular acceptance, is shown in Fig.~\ref{jetprops} (first panel) in the form $\rho_0(b) j^2(b)$. The dashed curve shows the result for a $4\pi$ acceptance.

 \subsection{Jet properties from jet correlations}

\begin{figure*}[h]
  \includegraphics[width=1.46in,height=1.65in]{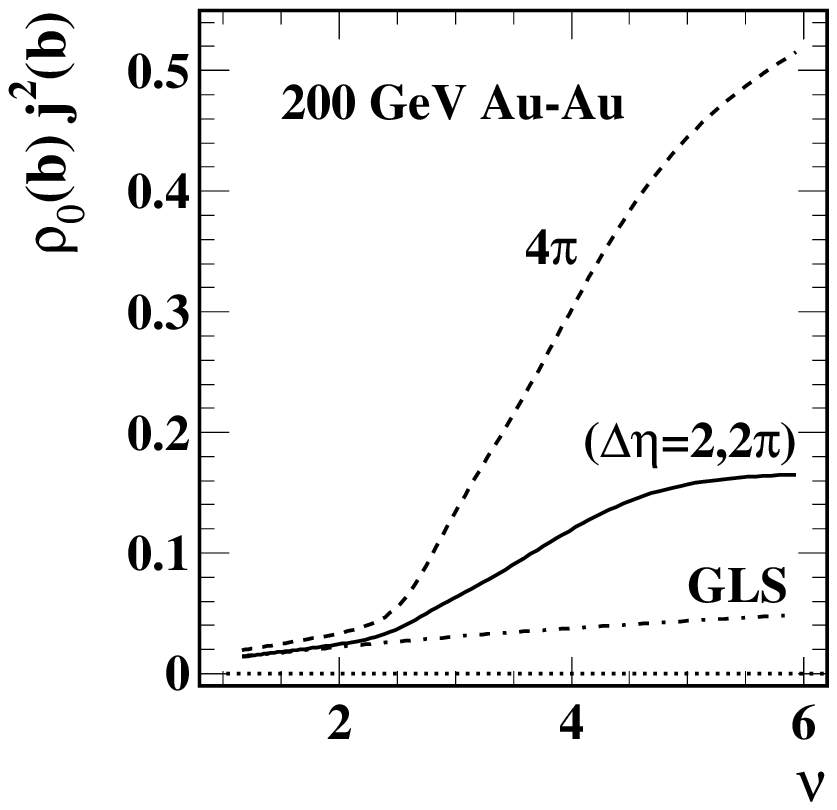}
   \includegraphics[width=1.46in,height=1.65in]{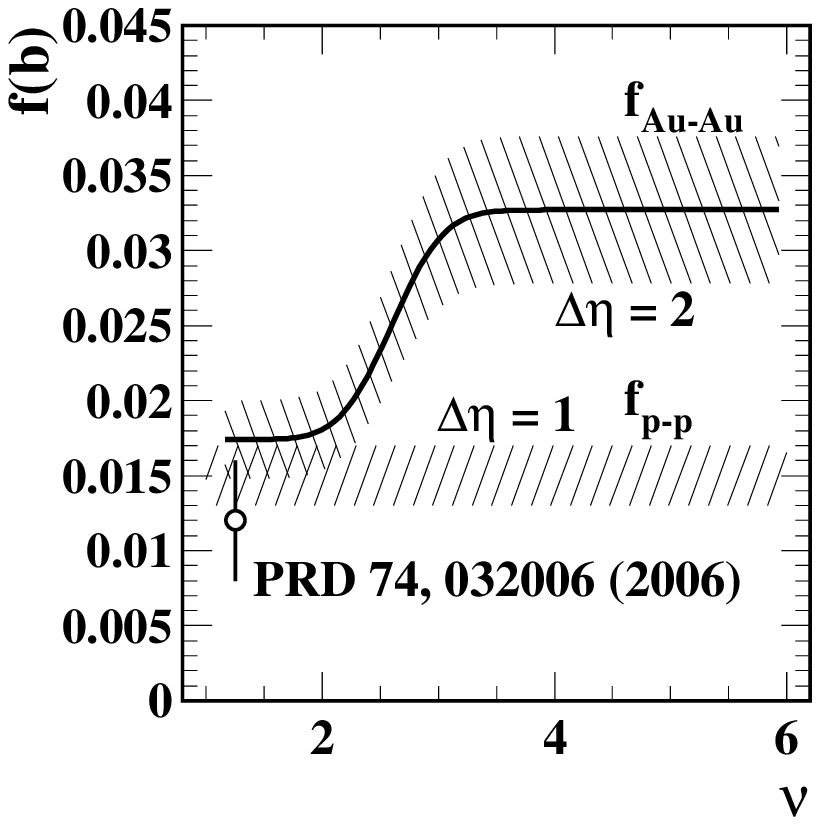}
  \includegraphics[width=1.46in,height=1.65in]{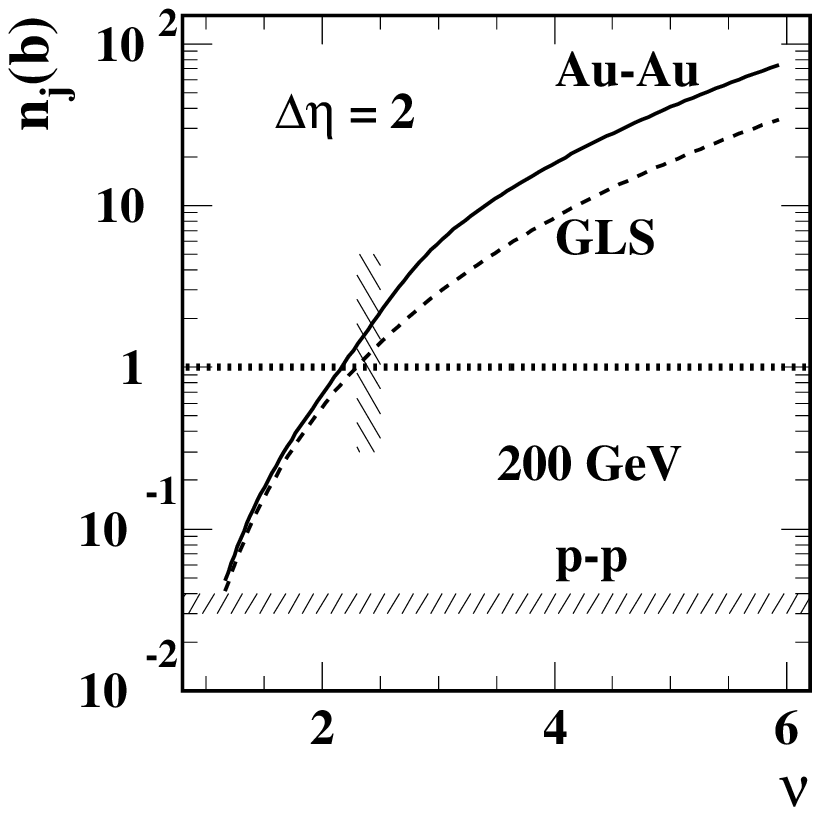}
 \includegraphics[width=1.46in,height=1.65in]{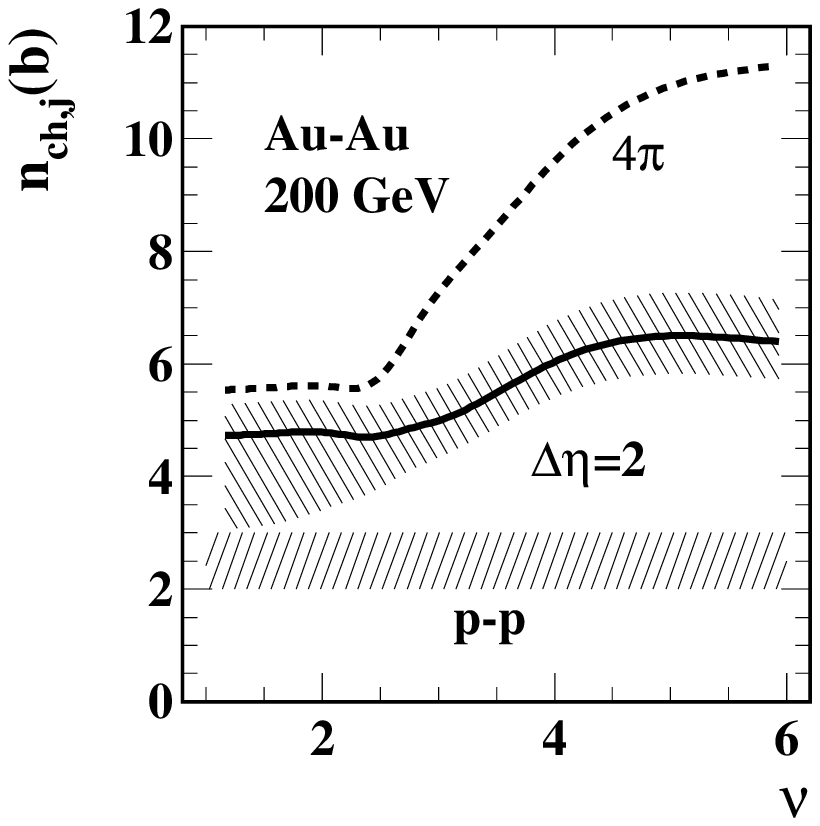}
\caption{\label{jetprops}
  First: Same-side 2D peak averaged over the angular acceptance (solid curve) and corresponding result for $4\pi$ acceptance (dashed curve).
Second: pQCD estimate of jet frequency $f(b)$ in Au-Au collisions (solid curve).
Third: Corresponding mean jet number per event vs Au-Au centrality (solid curve) and corresponding binary collision scaling (dashed curve).
Fourth: Mean per-jet fragment multiplicity inferred from trends in previous panels within the angular acceptance (lower curves) and in $4\pi$ (dashed curve).
}
 \end{figure*}

Fig.~\ref{jetprops} (second panel) shows a pQCD estimate of jet frequency $f(b)$ (solid curve) on A-A centrality measure $\nu$. The datum marked by the open symbol was inferred from a two-component analysis of spectra for 200 GeV p-p collisions with $\Delta \eta = 1$~\cite{ppprd}. The increase near $\nu = 2.5$ corresponds to the observation that the parton spectrum cutoff energy drops by about 10\% near the sharp transition, leading to an approximate 50\% increase in the dijet  cross section~\cite{fragevo}. 

Figure~\ref{jetprops} (third panel) shows the corresponding number of jets $n_j(b) = n_{bin}\, \Delta \eta\, f(b)$ for $\Delta \eta = 2$ within the STAR TPC. The sharp transition in jet correlation structure and spectra occurs at about the point in Au-Au collision centrality ($\nu \sim 2.5$, upper hatched region) where the number of jets becomes significantly greater than one. The dijet cross section is observed to increase by about 50\% at the same centrality. The dashed curve (GLS) is N-N binary collision scaling.

We now combine estimated jet frequencies with measured jet angular correlations to infer mean jet fragment multiplicities. 
Figure~\ref{jetprops} (fourth panel) shows the mean multiplicity derived from jet angular correlations. The assumption that the number of jet-correlated pairs is equal to the number of jets times the mean number of fragment pairs is expressed in the first line of
\bea \label{nchjj}
n_j(b)\, n_{ch,j}^2(b) 
&=& n_{ch}^2(b)\, j^2(b)  \\ \nonumber
 n_{ch,j}(b) &=& n_{ch}(b)\, \sqrt{j^2(b) / n_j(b)}.
\eea
The second line expresses the required factorization, where $n_{ch}(b) = 2\pi \Delta \eta\, \rho_0(b)$ (charged-particle multiplicity in the angular acceptance) and $j^2(b)$ are measured quantities. Mean per-jet fragment multiplicity $n_{ch,j}$ is thus inferred from correlation data and a pQCD jet number hypothesis. 
The fragment multiplicity for untriggered jets (mainly 3 GeV minijets) is something between 2 and 4 for p-p collisions, increasing to about 6 in central Au-Au, within the angular acceptance. The dashed curve is what the fragment multiplicity would be with $4\pi$ acceptance.  In more-central Au-Au collisions jets are elongated on $\eta$, and part of the jet falls outside the TPC angular acceptance.  Ironically, $n_{ch,j}$ is more uncertain in p-p than in central Au-Au collisions. The several curves and hatched regions illustrate the systematic uncertainty in the multiplicity estimate. We can now calculate the fragment density in Eq.~(\ref{fold2}) and the minimum-bias jet contribution to the final state.


 \subsection{Minijet (minimum-bias parton fragment) contribution to the final state}

\begin{figure*}[h] 
 \hfill  \includegraphics[width=1.75in,height=1.65in]{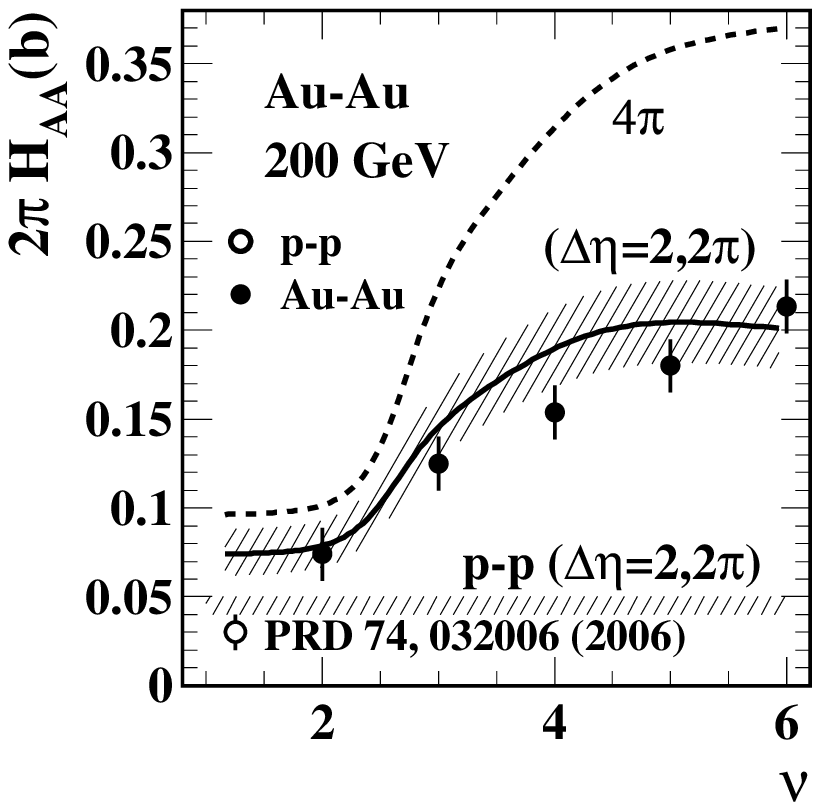} \hfill
   \includegraphics[width=1.75in,height=1.65in]{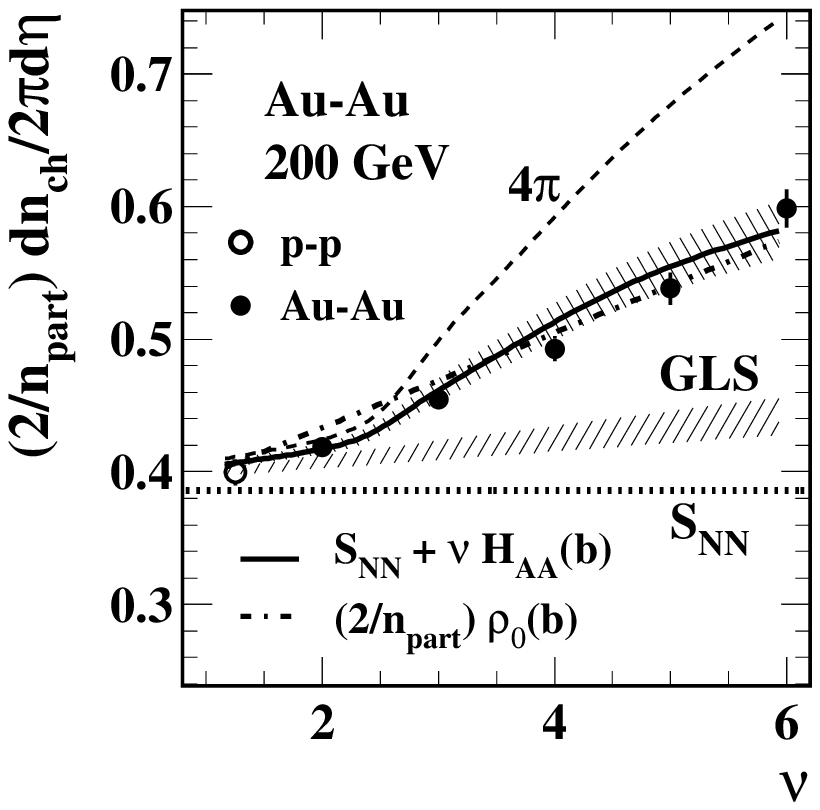} \hfill
    \includegraphics[width=1.75in,height=1.65in]{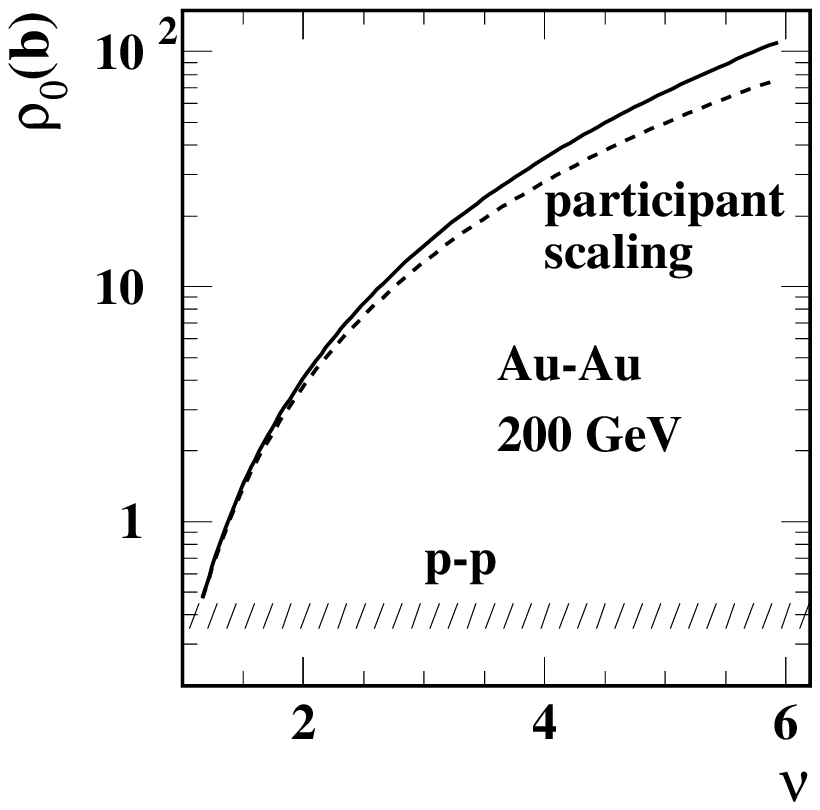} 
\hfill \hfill
\caption{\label{hardprod} First: Inferred $p_t$-integrated hard-component yield vs centrality. The open symbol is an estimate from Ref.~\cite{ppprd}.
Second: Total charged-particle yield vs centrality estimated from two-particle jet correlations (solid curve) and a two-component representation of measured single-particle data (dash-dotted line).
Third: Single-particle 2D angular density $\rho_0$ vs centrality. Participant scaling is indicated by the dashed curve.
 } 
 \end{figure*}

 Figure~\ref{hardprod} (first panel) shows spectrum hard component $H_{AA}(b)$ (solid curve) inferred from jet angular correlations according to Eq.~(\ref{fold2})
\bea \label{haaeq}
2\pi  H_{AA}(b) &=& \frac{dn_h}{d\eta} = f(b)\, n_{ch,j}(b).
\eea
The open point is an estimate from Ref.~\cite{ppprd}. The solid points are derived from the ``total hadrons'' data in Fig. 15 (left panel) of Ref.~\cite{hardspec}.
Multiplying through by $\nu / 2\pi$ gives the first line of
\bea  \label{fraction}
\nu H_{AA}(b) &=& \frac{2}{n_{part}}n_j(b)\frac{n_{ch,j}(b)}{2\pi \Delta \eta}\,  \\ \nonumber
&=&  \frac{2}{n_{part}} \rho_0(b)  \sqrt{n_j(b)\, j^2(b)}.
\eea
The second line incorporates the second line of Eq.~(\ref{nchjj}) and the definition of single-particle density $\rho_0(b)$. $\nu H_{AA}(b)$ is the hard component in the two-component spectrum model of Eq.~(\ref{aa2comp}). Figure~\ref{hardprod} (second panel) shows the two-component particle yield $S_{NN} + \nu H_{AA}(b)$ predicted by {\em measured} jet angular correlations (bold solid curve). Soft component $s_{NN}$ is by hypothesis fixed at $\sim 0.4$ [2D density on $(\eta,\phi)$] for all A-A centralities.  The solid points are the ``total hadrons'' data in Fig.~15 (left panel) of Ref.~\cite{hardspec} divided by $2\pi$.

The dash-dotted curve is the Kharzeev-Nardi (KN) approximation to the per-participant 2D density $(2/n_{part}) \rho_0(b)$ measured in more-central Au-Au collisions~\cite{nardi}. 
Figure~\ref{hardprod} (fourth panel) shows charged-hadron density $\rho_0(b)$, the 2D angular density on $(\eta,\phi)$ assumed for this analysis. 
The solid curve is the KN model $\rho_0(b) = (n_{part}/2) \rho_{pp}\{1+ x(\nu - 1)\}$ with $\rho_{pp} = 0.4$ and $x = 0.09$ for Au-Au at 200 GeV~\cite{nardi}.  
The KN description matches minimum-bias data in more-central collisions but fails in more-peripheral collisions where corrected yield data are sparse. For more-peripheral collisions we expect a GLS trend extrapolated from p-p collisions (dotted line in second panel). The sharp transition in jet angular correlations near $\nu = 2.5$ explains the deviation. 

Equation~(\ref{fraction}) (second line) implies that $\sqrt{n_j(b)\, j^2(b)}$ is the fractional hadron yield from parton fragmentation (minijet fractional yield). The first factor in the radicand is obtained from pQCD (relative systematic uncertainty $< 50$\%). The second factor is from measured jet angular correlations (relative uncertainty small). Combining angular correlation measurements and a pQCD estimate of jet number we find that one third of the hadronic final state in central Au-Au collisions at 200 GeV  is associated with {\em resolved jet correlations} (relative uncertainty $< 20$\%).


\section{Summary}


Hydro-motivated analysis of RHIC data tends to interpret the large hadron fragment contribution below 2 GeV/c in terms of flow phenomena. The role of parton scattering and fragmentation in nuclear collisions is minimized. Its pQCD description is artificially restricted to small regions of phase space. In contrast, model-independent analysis of spectrum and correlation structure reveals new fragmentation features quantitatively described by pQCD over the full momentum range.

Hard components extracted from $p_t$ spectra are now identified as single-particle manifestations of minimum-bias parton fragmentation in nuclear collisions. Spectrum hard components correspond quantitatively to minimum-bias jet angular correlations (minijets). pQCD fragment distributions  calculated by folding a minimum-bias parton energy spectrum with a parametrization of measured fragmentation functions accurately describe the measured hard components. 

Modification of fragmentation functions in more-central A-A collisions can be modeled by adjusting a single parameter in the FF parametrization (beta distribution) consistent with rescaling specific QCD splitting functions. The reference for all fragmentation in nuclear collisions is the FD derived from {\em measured} in-vacuum $e^+$-$e^-$ FFs and the parton spectrum for p-p collisions. 

Relative to the reference the spectrum hard component for p-p and peripheral Au-Au collisions is found to be {\em strongly suppressed} for smaller fragment momenta. At a specific point on centrality the Au-Au spectrum hard component transitions to strong enhancement at smaller momentum and suppression at larger momentum, described by FDs derived from medium-modified $e^+$-$e^-$ FFs.

Minimum-bias jet (minijet) correlations have been converted to absolute fragment yields  which are found to comprise approximately one third of the final state in central 200 GeV Au-Au collisions. Those results indicate that almost all large-angle scattered partons down to 3 GeV parton energy survive as jet manifestations in the final state, albeit with some modification.



Novel effects in A-A collisions may be related to strong color fields established among scattered energetic partons, an elaboration of three-jet events in LEP collisions.
Hadron fragment structure may directly reflect the large-scale color field geometry as a manifestation of local parton-hadron duality (LPHD).  
Newly-interpreted spectrum and correlation systematics, correctly associated with parton fragmentation, suggest evolution of the color-field geometry in nuclear collisions.
%
We conclude that pQCD calculations should be applied to all aspects of spectrum and correlation data in order to discover what is truly novel in RHIC collisions. We find that perturbative QCD {\em can} describe a large part of RHIC collision evolution -- and hydro interpretations are questionable.



\vskip .2in

Acknowledgments: I appreciate extensive discussions with and contributions from Lanny Ray, Duncan Prindle, David Kettler, Jeff Reid, Jeff Porter, Qingjun Liu, Dhammika Weerasundara, Aya Ishihara, Michael Daugherity, Hans Eggers and Rudy Hwa over the past fifteen years. This work was supported in part by the Office of Science of the US DOE under grant DE-FG03-97ER41020.


\end{document}